\documentclass{emulateapj}
\usepackage{apjfonts}

\slugcomment{Astrophys.J. 655 (2007) 233-260}
\shorttitle{Distances from Main Sequence Fitting. III}
\shortauthors{An et~al.}

\begin{document}

\title{The Distances to Open Clusters from Main-Sequence Fitting. III.\\
Improved Accuracy with Empirically Calibrated Isochrones}

\author{Deokkeun An\altaffilmark{1},
Donald M.\ Terndrup\altaffilmark{1},
Marc H.\ Pinsonneault\altaffilmark{1},
Diane B.\ Paulson\altaffilmark{2},\\
Robert B.\ Hanson\altaffilmark{3}, and
John R.\ Stauffer\altaffilmark{4}}

\altaffiltext{1}{Department of Astronomy, Ohio State University,
140 West 18th Avenue, Columbus, OH 43210;
deokkeun,terndrup,pinsono@astronomy.ohio-state.edu.}

\altaffiltext{2}{Planetary Systems Branch, Code 693,
NASA Goddard Space Flight Center, Greenbelt, MD 20771;
diane.b.paulson@gsfc.nasa.gov.}

\altaffiltext{3}{University of California Observatories/Lick
Observatory, Santa Cruz, CA 95064;
hanson@ucolick.org.}

\altaffiltext{4}{Infrared Processing and Analysis Center,
California Institute of Technology, Pasadena, CA 91125;
stauffer@ipac.caltech.edu.}

\begin{abstract}

We continue our series of papers on open cluster distances with a
critical assessment of the accuracy of main-sequence fitting using
isochrones that employ empirical corrections to the
color-temperature relations. We use four nearby open clusters with
multicolor photometry and accurate metallicities and present a
new metallicity for Praesepe (${\rm [Fe/H]} = +0.11 \pm 0.03$)
from high-resolution spectra. The internal precision of distance
estimates is about a factor of 5 better than the case without
the color calibrations. After taking into account
all major systematic errors, we obtain distances accurate to about
2\% -- 3\% when there exists a good metallicity estimate.
Metallicities accurate to better than 0.1 dex may be obtained from
$BVI_{C}K_{s}$ photometry alone. We also derive a helium abundance for
the Pleiades of $Y = 0.279 \pm 0.015$, which is equal within the
errors to the Sun's initial helium abundance and that of the Hyades.
Our best estimates of distances are $(m - M)_0 = 6.33 \pm 0.04$,
$8.03 \pm 0.04$, and $9.61 \pm 0.03$ to Praesepe, NGC~2516, and
M67, respectively. Our Pleiades distance at the spectroscopic
metallicity, $(m - M)_0 = 5.66 \pm 0.01$ (internal) $\pm 0.05$
(systematic), is in excellent agreement with several geometric
distance measurements. We have made calibrated isochrones for
$-0.3 \leq {\rm [Fe/H]} \leq +0.2$ available online.

\end{abstract}

\keywords{Hertzsprung-Russell diagram --- open clusters and
associations: individual (M67, NGC~2516, Pleiades, Praesepe) ---
stars: distances --- stars: abundances --- stars: evolution ---
stars: activity}

\section{INTRODUCTION}

The determination of accurate distances is the key to understanding how
stars and the Galaxy have formed and evolved. From protostars in
star-forming regions to ancient tracers of the halo, improved distances
have refined stellar evolutionary theory and Galactic structure models
\citep[e.g.,][]{reid99}. The {\it Hipparcos} mission \citep{perryman97b} was
especially valuable, providing trigonometric parallaxes for $\sim10^5$
stars to precision of 1--2 mas \citep{perryman97a}. These parallaxes,
however, are only useful for individual stars within $\sim 100$~pc. Most
open clusters are much more distant than this ``horizon,'' but
a half-dozen of the nearest clusters have 10 -- 50 or more {\it Hipparcos}
stars, yielding cluster parallaxes ostensibly accurate to 5\% or better
\citep{mermilliod97,perryman98,robichon99,vanleeuwen99}.

Main-sequence (MS) fitting, also known as the photometric parallax
method \citep[e.g.,][]{johnson57,siegel02}, has long been used to
estimate distances to individual stars and star clusters beyond
the limits of parallax studies, and is considered to be a robust
and well-understood technique. It was therefore a big surprise when
the {\it Hipparcos} distances to the Pleiades and Coma Ber open clusters
were in disagreement with distances from the MS-fitting at more than
a 3 $\sigma$ level \citep{pinsono98}. It is difficult to reconcile
a short Pleiades distance with stellar interior and spectroscopic
abundance studies. A high helium abundance would make a cluster fainter
than expected from its metallicity, and solutions of this type have
been discussed in the literature for the Pleiades \citep{belikov98}.
However, this is difficult to understand since there do not seem to be
nearby field stars of similar characteristics in the {\it Hipparcos} catalog
\citep{soderblom98}, and the helium enhancement would have to be
enormous ($Y \approx 0.34$). In addition, it has been suggested that
the metal abundance from spectroscopy may have been significantly
overestimated \citep{percival03}. An argument was also made that
distance estimates from theoretical stellar models have been
overestimated for young clusters due to unknown, age-related physics
\citep{vanleeuwen99}.

However, the most likely explanation is related to the {\it Hipparcos}
parallaxes themselves. \citet{pinsono98} showed that the 12 bright stars
near the center of the Pleiades all had virtually the same parallax,
$\sim 9$~mas, more than 1~mas larger than the mean parallax for
other cluster stars. They attributed this to a local zero-point
error of the individual stellar parallaxes that are correlated
over the {\it Hipparcos}' $0{\fdg}9$ field of view \citep{vanleeuwen98}.
These quasi-random errors were caused by the {\it Hipparcos} great-circle
data reductions, as \citet{makarov02,makarov03} proved by re-reducing
the Pleiades and Comar Ber cluster parallaxes in a different way that
correctly obtains the absolute zero point of parallax.
Additional effects may result from the way the {\it Hipparcos} data were
obtained and analyzed, and a more elaborate reduction of the {\it Hipparcos}
parallaxes promises to produce improved distances and better understood
errors \citep{vanleeuwen05a,vanleeuwen05b}.

The discrepant {\it Hipparcos} result for the Pleiades subsequently
led to many efforts to determine the cluster's distance from binaries
and independent parallax measurements
\citep[e.g.,][]{munari04,pan04,johnskrull05,soderblom05}.
These results support the longer distance scale from MS fitting,
verifying that the {\it Hipparcos} result was in error. With a formal
error of $\sim 1\%$ from these measurements, the Pleiades represents
a second system (besides the Hyades) with a sufficiently accurate
distance for a precision test of stellar evolutionary models.

Even though the controversy over the Pleiades distance is now settled,
a critical assessment of the MS fitting technique is still required to
reliably estimate a distance. In fact, MS fitting using theoretical
isochrones is a complex process that involves both physical and
empirical considerations \citep[e.g.,][]{stauffer01}. There
are, however, many opportunities to check the construction of the
isochrones. Stellar evolution models can be tested against the Sun
and other stars, such as eclipsing binaries, that have accurate
masses and radii.
Furthermore, multicolor photometry in nearby clusters and field
stars can be used to test the bolometric corrections and
color-effective temperature ($T_{\rm eff}$) relations to transform
theoretical quantities (luminosity and $T_{\rm eff}$) to magnitudes
and colors \citep[e.g.,][]{vandenberg03}.

In our first two papers of this series \citep[][hereafter Paper~I
and Paper~II, respectively]{pinsono03,pinsono04}, we began
a long-term effort to assess the accuracy of distances from MS fitting
and to reduce or eliminate systematic errors in the process,
particularly those arising from the transformation of theoretical
to observational quantities. In Paper~I, we demonstrated that stellar
models from the Yale Rotating Evolutionary Code \citep[YREC;][]{sills00}
are in good agreement with masses and luminosities for the
well-studied Hyades eclipsing binary vB~22 \citep{torres02}.
These models also satisfy stringent tests from helioseismology,
and predict solar neutrino fluxes in line with observations
\citep{basu00,bahcall01,bahcall04}.
In Paper~II, we showed that the models provide a good match to the
spectroscopically determined temperatures \citep{paulson03} for
individual Hyades members with good parallaxes \citep{debruijne01}.
However, we found that any of the widely-used color-$T_{\rm eff}$
relations \citep[e.g.,][]{alonso95,alonso96,lejeune97,lejeune98} fail
to reproduce the observed shapes of the MS in the Hyades; differences
in broadband colors were as large as $\sim0.1$ mag. The existence
of these systematic errors in the colors in the presence of agreement
between the spectroscopic and theoretical $L-T_{\rm eff}$ scales
strongly implies that there are problems with the adopted
color-$T_{\rm eff}$ relations instead of errors in the theoretical
$T_{\rm eff}$ scale. Therefore, we proposed empirical corrections
to the color-$T_{\rm eff}$ relations from \citet{lejeune97,lejeune98}
that were adopted in the isochrone computations.

In this study, we generate a set of isochrones over a wide range
of age and metallicity, and test the validity of the Hyades-based
color-$T_{\rm eff}$ corrections using extensive multicolor photometry of
four well-studied nearby open clusters. We show that isochrones employing
the Hyades empirical corrections precisely match the observed MS
shapes, except where anomalously blue colors in young open clusters
have been previously noted \citep{stauffer03}. Furthermore, we
demonstrate that the empirical corrections improve the internal
precision of the isochrones by examining the consistency of distances
derived from several color-magnitude diagrams (CMDs).

We also assess various sources of systematic errors in the MS-fitting
technique. Previously, \citet{pinsono98} considered the effects of age,
metal abundance, helium, reddening, and systematic errors in
the photometry, demonstrating that these could not explain the short
distance to the Pleiades from {\it Hipparcos}. \citet{terndrup02} paid
attention to the adopted reddening law in a discussion of the distance
to NGC~2516. Here we extend the error analysis more quantitatively,
emphasizing photometric calibration issues and the bias in distance
estimates induced by the presence of unresolved cluster binaries or
field foreground/background stars.

This paper also explores the effect of metallicity on the luminosity
of the MS. Metallicity changes isochrone luminosities more strongly
than many other input parameters, and the degree of sensitivity
depends on the color index used. This permits a purely photometric
derivation of the metallicity
\citep[e.g.,][]{pinsono98,pinsono00,stello01,terndrup02},
which can be compared to metallicities derived from high-resolution
spectra. An agreement between the photometric and spectroscopic
metallicities, as we find in this paper, provides supporting
evidence that the effects of metallicity on the theoretical
quantities ($L$, $T_{\rm eff}$) and on the color-$T_{\rm eff}$
relations are correctly computed.

The distances in this paper are tied to the Hyades distance
at $(m - M)_0 = 3.33 \pm 0.01$ ($d = 46.34 \pm 0.27$ pc),
the cluster's center-of-mass inferred from the {\it Hipparcos}
catalog \citep{perryman98}. Unlike the controversial {\it Hipparcos}
distance to the Pleiades, the large angular diameter of the Hyades on
the sky makes the cluster parallax less vulnerable to the spatial
correlation of the {\it Hipparcos} parallax
\citep{narayanan99a,narayanan99b,debruijne01}.

In \S~2 we compile cluster photometry, metallicities, reddening
estimates, and information on binarity and membership, and present
a metallicity for Praesepe from new high-resolution spectra. In \S~3
we briefly describe the construction of the isochrones. In \S~4 we
compute the distances to the sample clusters with the reddening fixed
at previously known values and demonstrate that the empirical
corrections improve the internal precision of the isochrones. In \S~5
we simultaneously solve for the cluster metallicity, reddening, and
distance from the $\chi^2$ minimization. In \S~6 we evaluate
the effects of several systematic error sources, including those from
cluster binaries and field star contamination. In \S~7 we discuss
several implications of our results. In the Appendix we address
issues on the photometric zero points of the empirical Hyades isochrone.

\section{OPEN CLUSTER DATA}

\subsection{Selection of Clusters}\label{sec:sample}

We consider four nearby Galactic open clusters with extensive
multicolor photometry: Praesepe (= M44; NGC~2632), the Pleiades (= M45),
M67 (= NGC~2682), and NGC~2516.
The choice of these clusters was motivated by several factors. All
have well-determined estimates of metal abundance and reddening
against which we can compare photometrically-derived values. Samples
in Praesepe, the Pleiades, and M67 are dominated by known
cluster members, so we can examine whether the Hyades-based color
calibration from Paper~II generates isochrones that precisely
match the shapes of the MS in these clusters.
Praesepe has extensive information on binarity, so systematic
errors in distances arising from the presence of binaries can be
explored. M67 and NGC~2516 each have modern photometry from two
independent studies, from which we gauge the sizes of errors that arise
from photometric calibration issues. In addition,
NGC~2516 has a relatively high
reddening compared to the other clusters, which allows us to
evaluate the consequences of adopting particular reddening laws.

\begin{deluxetable*}{llc}
\tablewidth{0pt}
\tablecaption{Recent Measurements of the Pleiades Distance\label{tab:hipp}}
\tablehead{
  \colhead{Reference} &
  \colhead{Method} &
  \colhead{$(m -  M)_0$}
}
\startdata
\citet{narayanan99b} & Moving cluster              & $5.58 \pm 0.18$ \nl
\citet{gatewood00}   & Ground-parallax             & $5.58 \pm 0.12$ \nl
\citet{makarov02}    & Hipparcos reanalysis        & $5.55 \pm 0.06$ \nl
\citet{munari04}     & Eclipsing binary (HD~23642) & $5.60 \pm 0.03$ \nl
\citet{pan04}        & Astrometric binary (Atlas)  & $5.65 \pm 0.03$ \nl
\citet{zwahlen04}    & Astrometric binary (Atlas)  & $5.60 \pm 0.07$ \nl
\citet{johnskrull05} & {\it HST} parallax          & $5.66 \pm 0.06$ \nl
\citet{soderblom05}  & {\it HST} parallax          & $5.65 \pm 0.05$ \nl
\citet{southworth05} & Eclipsing binary (HD~23642) & $5.72 \pm 0.05$ \nl
Weighted mean        & \nodata                     & $5.63 \pm 0.02$
\enddata
\end{deluxetable*}

The Pleiades is a special case because its distance has recently
been accurately measured from astrometric and eclipsing binary
studies and from ground- and space-based parallaxes, allowing
a precise test of distances derived from MS fitting. Individual
measurements of these studies are summarized in Table~\ref{tab:hipp},
and the weighted average distance from these measurements is
$(m - M)_0 = 5.632 \pm 0.017$. The {\it Hipparcos} distances to
the Pleiades and other clusters are discussed in \S~\ref{sec:hipp}.

We use cluster ages from \citet{meynet93} for the Pleiades (100 Myr),
M67 (4 Gyr), and NGC~2516 (140 Myr). Praesepe is generally considered
to be the same age as the Hyades \citep[e.g.,][]{mermilliod81b};
in Paper~II, we assumed an age of 550 Myr for the Hyades, as derived
from isochrones without overshooting \citep{perryman98}. As we
demonstrate, the cluster distances are insensitive to the choice of age.

\subsection{Photometry}\label{sec:photometry}

\subsubsection{Praesepe and the Pleiades}

We compiled optical photometry for Praesepe and the Pleiades mainly
from WEBDA \citep{mermilliod03}\footnote{See http://obswww.unige.ch/webda/webda.html.}
and the Open Cluster
Database.\footnote{See http://spider.ipac.caltech.edu/staff/stauffer/opencl/index.html.}
Data in $V$ and $B - V$ for Praesepe came from \citet{johnson52},
\citet{dickens68}, and \citet{castelaz91}. Following the suggestion
by \citet{dickens68}, we added $+0.002$ to $V$ and $+0.006$ to their
$B - V$ to match the \citet{johnson52} data. We adopted the photometry from
\citet{castelaz91} without alteration; the average difference in $B - V$
is $+0.002 \pm 0.012$, \citet{johnson52} being redder, among 39 stars
in common (including nonmembers). We compared the numerous sources of
$V$ and $B - V$ in the Pleiades, but did not find statistically
significant differences in any sample. At $B - V \ga 0.8$, the scatter
in measurements from several sources often exceeded the stated
photometric errors; this may result from brightness and color changes
from stellar spots on rapidly rotating stars in young open clusters
\citep{stauffer03}. For stars in this color range, as everywhere
else, we simply averaged the several available colors and magnitudes.

The situation in $V - I_C$ is less straightforward because data for
Praesepe and the Pleiades, like those for the Hyades, came from either
the Johnson ($V - I_J$) or the Kron ($V - I_K$) systems.
The colors on the Johnson system were transformed to the
Cousins ($V - I_C$) system using an updated transformation equation
as described in the Appendix. The Kron colors were transformed to
the Cousins system using the cubic polynomial derived by \citet{bessell87}.

In Praesepe, stars with $V < 12$ have $V - I$ photometry on the Johnson
system from \citet{mendoza67} and \citet{castelaz91}, while fainter stars
have photometry on the Kron system from \citet{upgren79},
\citet{weis81}, \citet{stauffer82b}, and \citet{mermilliod90}.
Intercomparisons showed that the \citet{mendoza67} and
\citet{castelaz91} colors were on the same system. The same is true for
the Kron data except that the photometry from \citet{upgren79}
required a redward shift of $+0.03$ in $V - I_K$ to match the other photometry.
In the Pleiades, stars with $V < 10$ have $V - I$ photometry on the
Johnson system from \citet{johnson66}, \citet{mendoza67}, and
\citet{iriarte69}, while fainter stars have photometry on the Kron system
from \citet{stauffer82a}, \citet{stauffer82c}, \citet{stauffer84},
\citet{stauffer87}, \citet{stauffer89}, \citet{prosser91}, and
\citet{stauffer94}. Direct comparison between the \citet{mendoza67}
photometry and that of \citet{iriarte69} showed that they agree well
for $V - I_J \leq 0.5$, but the \citet{mendoza67} data are systematically
redder by 0.02 mag for stars with $V - I_J \geq 0.5$. We opted to adjust
the red \citet{mendoza67} photometry to place them on
the \citet{iriarte69} scale.

\subsubsection{M67 and NGC~2516}

\begin{deluxetable*}{llccc}
\tablewidth{0pt}
\tablecaption{Differences in the Photometry\label{tab:phot.diff}}
\tablehead{
\colhead{Cluster} &
\colhead{Comparison} &
\colhead{$\langle \Delta V \rangle$} &
\colhead{$\langle \Delta (B -  V)  \rangle$} &
\colhead{$\langle \Delta (V -  I)_C \rangle$}
}
\startdata
M67      & MMJ93 -- S04    & $+0.017\pm0.002$ & $+0.009\pm0.002$ & $+0.022\pm0.001$ \nl
NGC~2516 & S02 -- JTH01    & $+0.016\pm0.001$ & $-0.003\pm0.001$ & $+0.011\pm0.001$ \nl
Hyades   & Ground -- Tycho-1\tablenotemark{a} & $+0.012\pm0.002$ & $-0.009\pm0.001$ & \nodata \nl
Praesepe & Ground -- Tycho-1\tablenotemark{a} & $+0.009\pm0.007$ & $+0.004\pm0.003$ & \nodata \nl
Pleiades & Ground -- Tycho-1\tablenotemark{a} & $-0.017\pm0.003$ & $+0.000\pm0.002$ & \nodata \nl
r.m.s.   & \nodata         & $0.015$          & $0.006$          & $0.017$
\enddata
\tablenotetext{a}{Computed for stars with $V_T \leq 9$.}
\end{deluxetable*}

For M67, we used $BVI_C$ photometry from
\citet[][hereafter MMJ93]{montgomery93} and also from
\citet[][hereafter S04]{sandquist04}; these are analyzed separately.
S04 provided a comparison between the two, revealing statistically
significant differences between the two studies. In
Table~\ref{tab:phot.diff} we compile the mean differences in the
photometry for M67 and for other clusters as is discussed
below. The first column of the table lists the cluster name, and
the sense of the comparison is shown in the second column. The
last three columns display the mean difference and its standard error
in $V$, $B - V$, and $V - I_C$, respectively. In comparison to S04,
the MMJ93 data are fainter in $V$ and are redder in $V - I_C$. As shown
in Figure~5 of S04, the differences are largest for those stars with
$V - I_C \la 1.0$.

\begin{figure}
\epsscale{1.2}
\plotone{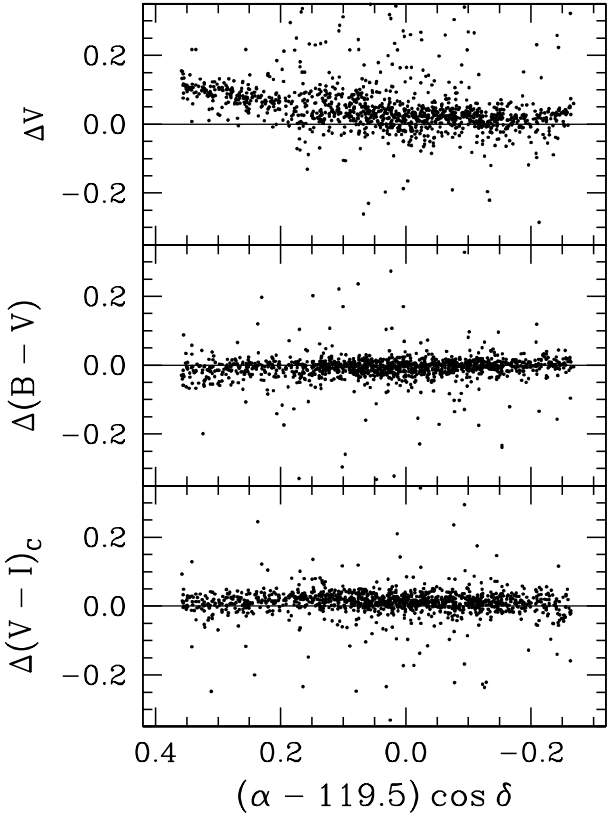}
\caption{Comparison of NGC~2516 photometry as a function of right
ascension (in degrees). The differences are in the sense of the S02
minus the JTH01 photometry.\label{fig:25.compra}}
\end{figure}

\begin{figure}
\epsscale{1.2}
\plotone{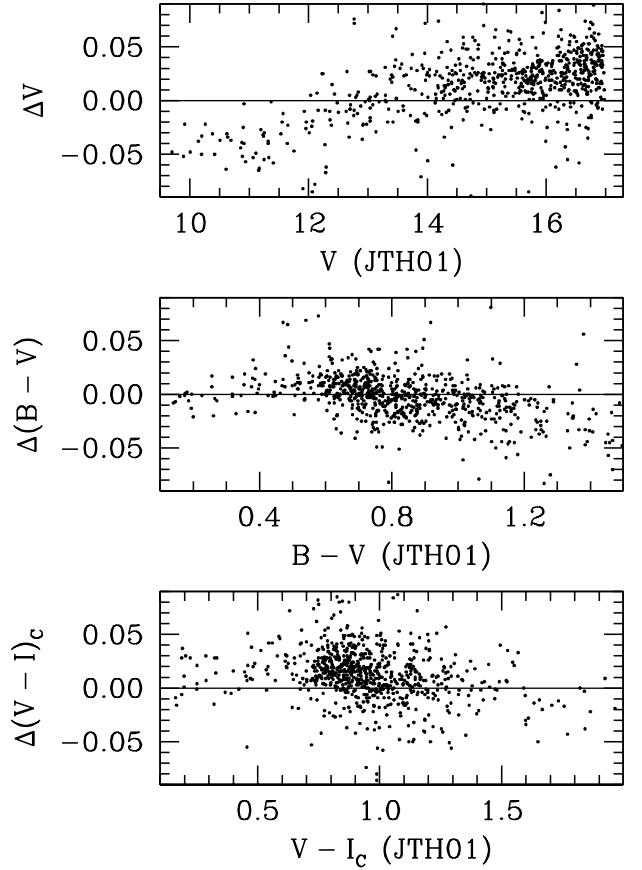}
\caption{Comparison of NGC~2516 photometry,
excluding stars observed under non-photometric conditions by
S02. The differences are in the sense of the S02 minus the
JTH01 photometry.\label{fig:25.compv}}
\end{figure}

For NGC~2516, we have independent photometry in $BVI_C$ from
\citet[][hereafter JTH01]{jeffries01} and from \citet[][hereafter
S02]{sung02}. Neither study compares their photometry with the
other. In Figure~\ref{fig:25.compra} we plot the differences in
the photometry from the two studies against right ascension, which
reveals a position-dependent difference in $V$ (but not in the
colors). The sense of the difference is that the photometry
towards the east is fainter in the S02 study. According to
S02, photometry in this portion of the sky was obtained on a
non-photometric night which was then adjusted to match the data
in the rest of their survey using stars also observed under good
conditions. As their paper lists only the average values, we chose
to use the photometry only from regions obtained entirely under
photometric conditions. However, there still remain significant
differences, as shown in Figure~\ref{fig:25.compv}, between the
JTH01 and the S02 photometry even when the eastern data in the
latter study are removed. Compared to the JTH01 values, the S02
data are brighter in $V$ and redder in $V - I_C$ at the top of the
MS, but are fainter in $V$ and bluer in $B - V$ at the faint end.
The mean differences are summarized in Table~\ref{tab:phot.diff}.

\subsubsection{Assignment of Random Errors}

Most of the collated photometry lacks errors for individual stars,
and those errors reported by MMJ93, JTH01, and S02 are typically
from a small number of repeat observations. Our MS-fitting procedure
(\S~\ref{sec:filtering}) first removes stars that are statistically
separated from the MS, and this in turn requires a suitable error
for each star.

In Praesepe and the Pleiades, we divided the data into bins in
$V$, each 2 mag wide, and then computed the median of
rms in magnitude and colors for the relatively few stars with
multiple measurements. We assigned this median value to all data points
in each bin as random photometric errors. For NGC~2516, we followed
the same binning procedure, then removed a systematic trend in
the differences between the JTH01 and S02 studies by subtracting
a linear function in $V$. We then computed the rms of
the differences in $V$, $B - V$, and $V - I_C$ in each bin, divided
these by $\sqrt{2}$, and assigned these errors to both data sets
assuming that each study would have about
the same errors. For M67, we followed the same procedure, first
matching the stars by their coordinates, and computing the rms
of the differences after removing a systematic trend in $V$. S04
reports the error for each star based on a large number of measurements.
We used those errors directly in our analysis of the S04 data, and
assumed that the scatter between the two studies was dominated by
measurement errors in MMJ93. We set the errors for the MMJ93 study
from the rms of the differences with respect to the S04 data.

\subsubsection{2MASS $K_s$}

We calculated $V - K_s$ colors from the All Sky Data Release of
the Two Micron All Sky Survey (2MASS) Point Source Catalog
(PSC).\footnote{See http://www.ipac.caltech.edu/2mass/.} Here $K_s$
designates the ``short''-$K$ filter in 2MASS \citep{carpenter01}.
Based on PSC flag parameters, we excluded stars that were saturated
or undetected. We also ignored blended or contaminated sources.
The $V - K_s$ errors were taken as the quadrature sum of $V$ errors
and the catalog's ``total'' photometric uncertainties in $K_s$.

\subsection{Systematic Errors in the Photometry}

Our analysis of systematic errors in the MS-fitting method
(\S~\ref{sec:sys}) shows that calibration errors in the photometry
contribute significantly to the overall error budget. In addition
to the direct comparisons between the studies of M67 and
NGC~2516, we can use the Tycho-1 photometry \citep{vanleeuwen97}
to check whether photometry of the Hyades, Praesepe, and the Pleiades
is on a consistent system (the other clusters are too distant to
have many stars in the Tycho catalog).
Here we assume that the Tycho photometry is consistently on the same
scale in all parts of the sky. In Table~\ref{tab:phot.diff}, we
display the mean differences between the ground-based photometry for
these clusters and the Tycho values. The latter were transformed to
the Johnson system using the equations in \citet{oja98} for $V$ and
in \citet{mamajek02} for $B - V$. The values in the table were computed
for stars with $V_T \leq 9$.

\begin{figure*}
\epsscale{0.9}
\plotone{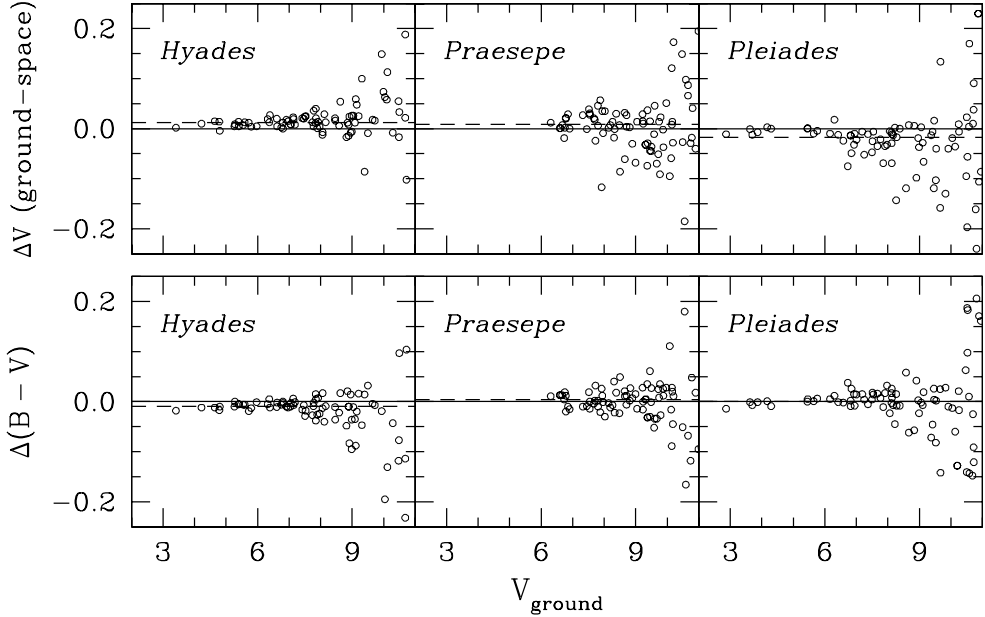}
\caption{Differences in $V$ ({\it top}) and $B - V$ ({\it bottom})
between the ground-based and the Tycho-1 photometry, after transforming
the latter into the Johnson system (see text). The differences are in
the sense of the ground minus the Tycho-1 values. The solid line indicates
equality, while the dashed line is plotted at the weighted mean
difference for stars with $V_T \leq 9$ (Table~\ref{tab:phot.diff}).
The scatter for $V > 8$ mainly reflects errors in the Tycho photometry.
\label{fig:tycho}}
\end{figure*}

The comparison between the Tycho-1 values and the ground-based
data is shown in Figure~\ref{fig:tycho}. The scatter is usually
dominated by errors in the ground-based photometry for $V \leq 8$
and by the Tycho photometry for fainter stars. The solid line in
each panel denotes equality, while the dashed line shows the mean
difference from Table~\ref{tab:phot.diff}.

\begin{deluxetable}{llcc}
\tablewidth{0pt}
\tablecaption{Differences in the Tycho Photometry\label{tab:tycho.diff}}
\tablehead{
\colhead{Cluster} &
\colhead{Comparison} &
\colhead{$\langle \Delta B_T \rangle$} &
\colhead{$\langle \Delta V_T \rangle$}
}
\startdata
Hyades   & Tycho-2 -- Tycho-1 & $-0.009\pm0.002$ & $-0.006\pm0.002$ \nl
Praesepe & Tycho-2 -- Tycho-1 & $-0.018\pm0.004$ & $-0.025\pm0.004$ \nl
Pleiades & Tycho-2 -- Tycho-1 & $-0.014\pm0.003$ & $-0.016\pm0.003$
\enddata
\tablecomments{Computed for stars with $V_T \leq 9$.}
\end{deluxetable}

In Table~\ref{tab:tycho.diff} we list the differences between the
Tycho-1 and Tycho-2 data \citep{hog00} for bright stars in
the Hyades, Praesepe, and the Pleiades. The differences were computed
directly with the Tycho values (again for $V_T \leq 9$), without
transformations to the Johnson system.
Taken together, the data in Tables~\ref{tab:phot.diff} and
\ref{tab:tycho.diff} show that the various studies probably could
have calibration errors on the order of 0.01 -- 0.02 mag in $V$, but
smaller in $B - V$. The situation in $V - I_C$ is less well
determined, mainly because we have fewer studies to compare.

In Paper~II we also noted that the Tycho-1 values for $V$ were
brighter than the ground-based data by $+0.012$; we opted to add
this constant to the Tycho data before averaging with the
ground-based photometry. Had we chosen to adopt the space-based
scale instead, our calibrated isochrones would be brighter in $V$
and also bluer in $V - K_s$ since the $V$ and $K_s$ data were
obtained independently. This correction was not applied to the
isochrones used in this paper. In the Appendix we discuss other
issues involved in the isochrone calibration.

\subsection{Membership and Binarity}\label{sec:membin}

We also collated information on binarity and membership in
Praesepe and the Pleiades from WEBDA and the Open Cluster Database.
Any star that was listed as a nonmember for any reason (e.g.,
from photometry or radial velocities) was rejected. In Praesepe,
there are extensive data on binarity \citep[e.g.,][]{mermilliod99,bouvier01}.
Stars that were designated as likely or probable binary stars were
flagged. In the Pleiades, a considerable number of stars have binarity
and membership information \citep{jones81,stauffer91,mermilliod92,
schilbach95,bouvier97,belikov98,moraux01}. We rejected any star with
a membership probability less than 50\% but kept any star that did
not have a membership probability. S04 provides
a list of single cluster members based on proper motion surveys and
known binary systems in the literature and further elimination
of unrecognized binaries in CMDs from his high-precision photometry.
We used this information for both MMJ93 and S04 photometry
in this paper, unless otherwise specified.

\subsection{Metallicity and Reddening}

\subsubsection{New Spectroscopy in Praesepe}\label{sec:spec}

Obtaining accurate distances relative to our calibrating cluster,
the Hyades, requires accurate relative metallicities because
the luminosity of the MS is sensitive to the metal abundance. As part
of this study, we determined a new metallicity for Praesepe using
the same method used to find the Hyades abundance \citep{paulson03}.
The relative abundance of Praesepe with respect to the Hyades
should therefore be accurate since systematic errors arising from
the solar abundances, employed model atmospheres, oscillator strengths,
or effective temperature scales would be minimized.

We obtained spectra of four Praesepe stars with the Magellan
Inamori Kyocera Echelle (MIKE) spectrograph \citep{bernstein03} on
the Magellan 6.5 m Clay telescope at Las Campanas Observatory.
A 0.35\arcsec wide slit gave a resolving power of $\sim55,000$ per
resolution element (4 pixels) and wavelength coverage from
4500 to 9200\thinspace \AA. The spectra are of high
quality, with S/N $> 100$.

\begin{deluxetable}{lcccccc}
\tablewidth{0pt}
\tablecaption{Spectroscopy in Praesepe\label{tab:spec}}
\tablehead{
  \colhead{} &
  \colhead{$T_{\rm eff}$} &
  \colhead{$\log{g}$} &
  \colhead{$\xi$} &
  \colhead{} &
  \colhead{} &
  \colhead{} \nl
  \colhead{ID\tablenotemark{a}} &
  \colhead{\rm (K)} &
  \colhead{${\rm (cm\ s^{-2})}$} &
  \colhead{${\rm (km\ s^{-1})}$} &
  \colhead{$\log \epsilon$} &
  \colhead{$\sigma_{\log \epsilon}$} &
  \colhead{\rm [Fe/H]}
}
\startdata
KW~23 &5700 &4.5 &0.5 &7.78 &0.05 &0.10 \nl
KW~58 &5850 &4.5 &0.5 &7.80 &0.06 &0.12 \nl
KW~304&5625 &4.4 &0.4 &7.80 &0.06 &0.12 \nl
KW~336&5600 &4.5 &0.4 &7.79 &0.06 &0.11
\enddata
\tablenotetext{a}{KW = Klein Wassink (1927).}
\end{deluxetable}

The spectra were reduced using standard IRAF\footnote{IRAF is distributed
by the National Optical Astronomy Observatory, which is operated by
the Association of Universities for Research in Astronomy, Inc., under
cooperative agreement with the National Science Foundation.} packages.
The stellar parameters -- $T_{\rm eff}$, surface gravity ($\log{g}$),
microturbulence ($\xi$), and {\rm [Fe/H]} -- were derived following the
procedure in \citet{paulson03}, and are listed in Table~\ref{tab:spec}.
We employed the spectral synthesis code MOOG \citep{sneden73} and used
stellar model atmospheres based on the 1995 version of the ATLAS9 code
\citep{castelli97}. Within IRAF, we used Gaussian fits to 15
\ion{Fe}{1} lines and nine \ion{Fe}{2} lines, a subsample of those listed in
Table~1 of \citet{paulson03}. The $T_{\rm eff}$ was derived by requiring
that individual line abundances be independent of excitation potential
and that microturbulence ($\xi$) be independent of line strength.
Insisting on ionization equilibrium between \ion{Fe}{1} and \ion{Fe}{2}
allowed for the simultaneous determination of $\log{g}$ with
$T_{\rm eff}$ and microturbulence ($\xi$). Errors in $T_{\rm eff}$
are $\sim50$~K. We analyzed a reflected light spectrum of the asteroid
Iris in order to obtain an instrumental correction to the solar
$\log \epsilon$(Fe). Using this correction, we obtained
${\rm [Fe/H]} = 0.11 \pm 0.03$ (s.e.m.\footnote{Standard error
of the mean.}) for the Praesepe cluster.

\subsubsection{Metallicity and Reddening Estimates in the Literature}

\begin{deluxetable}{lcc}
\tablewidth{0pt}
\tablecaption{Cluster Metallicity from High-Resolution Spectroscopy
\label{tab:feh.lit}}
\tablehead{
  \colhead{Reference} &
  \colhead{{\rm [Fe/H]}} &
  \colhead{s.e.m.}
}
\startdata
\multicolumn{3}{c}{Praesepe} \nl
\hline
Boesgaard \& Friel\tablenotemark{a} & $+0.09$ & $0.03$ \nl
This paper (adopted value)          & $+0.11$ & $0.03$ \nl
\hline
\multicolumn{3}{c}{Pleiades} \nl
\hline
\citet{cayrel88}                    & $+0.13$ & $0.07$ \nl
Boesgaard \& Friel\tablenotemark{b} & $+0.03$ & $0.02$ \nl
\citet{king00}                      & $+0.06$ & $0.05$ \nl
Weighted mean                       & $+0.04$ & $0.02$ \nl
                                    & \nodata & ($0.03$)\tablenotemark{c} \nl
\hline
\multicolumn{3}{c}{M67} \nl
\hline
\citet{lopez88}          & $+0.04$ & $0.04$ \nl
\citet{hobbs91}          & $-0.06$ & $0.03$ \nl
\citet{friel92}          & $+0.02$ & $0.12$ \nl
\citet{tautvaisiene00}   & $-0.03$ & $0.01$ \nl
\citet{yong05}           & $+0.02$ & $0.02$ \nl
\citet{randich06}        & $+0.03$ & $0.01$ \nl
Weighted mean            & $+0.00$ & $0.01$ \nl
                         & \nodata & ($0.03$)\tablenotemark{c} \nl
\hline
\multicolumn{3}{c}{NGC~2516} \nl
\hline
\citet{terndrup02}     & $+0.01$ & $0.07$
\enddata
\tablenotetext{a}{From \citet{boesgaard89} and \citet{friel92} after excluding known nonmembers.}
\tablenotetext{b}{From \citet{boesgaard89} and \citet{boesgaard90} after excluding known nonmembers.}
\tablenotetext{c}{Standard deviation of the measurements.}
\end{deluxetable}

For other clusters in this study, we adopted or calculated average
metallicities from high-resolution spectroscopy as summarized in
Table~\ref{tab:feh.lit}. For the studies by Boesgaard \& Friel in
Praesepe and the Pleiades, we have recomputed the cluster
averages from individual {\rm [Fe/H]} estimates after excluding
cluster nonmembers. The only existing measurement of the abundance
of NGC~2516 using high-resolution spectroscopy is from
\citet{terndrup02} which was derived from only two stars.

\begin{deluxetable*}{llll}
\tablewidth{0pt}
\tablecaption{Cluster Reddening in the Literature\label{tab:ebv.lit}}
\tablehead{
  \colhead{Reference} &
  \colhead{Method} &
  \colhead{$E(B -  V)$} &
  \colhead{s.e.m.}
}
\startdata
\multicolumn{3}{c}{Praesepe}\nl
\hline
\citet{mermilliod81a}            & $UBV$                            & $0.00 $ & $0.01$\tablenotemark{a}\nl
\citet{nicolet81}                & Geneva                           & $0.011$ & $0.005$\nl
\citet{nissen88}                 & $uvby$--$\beta$\tablenotemark{b} & $0.007$ & $0.002$\nl
Weighted mean                    & \nodata                          & $0.007$ & $0.002$\nl
                                 & \nodata                          & \nodata & ($0.003$)\tablenotemark{c}\nl
\hline
\multicolumn{3}{c}{Pleiades}\nl
\hline
\citet{mermilliod81a}            & $UBV$                                     & $0.04 $ & $0.01$\tablenotemark{a}\nl
\citet{nicolet81}                & Geneva                                    & $0.062$ & $0.005$\nl
\citet{breger86}\tablenotemark{d}& Sp-type, $uvby$--$\beta$\tablenotemark{b} & $0.044$ & $ 0.003$\nl
\citet{nissen88}                 & $uvby$--$\beta$\tablenotemark{b}          & $0.039$ & $0.005$\nl
Weighted mean                    & \nodata                                   & $0.046$ & $0.005$\nl
                                 & \nodata                                   & \nodata & ($0.009$)\tablenotemark{c}\nl
\citet{breger86} [adopted value]\tablenotemark{e} & Sp-type, $uvby$--$\beta$\tablenotemark{b} & $0.032$ & $ 0.003$\nl
\hline
\multicolumn{3}{c}{M67}\nl
\hline
\citet{taylor80}                 & Multiple approaches            & $0.046$ & $0.006$ \nl
\citet{eggen81}                  & $uvby$--$\beta$\tablenotemark{b} & $0.050$ & $0.013$ \nl
\citet{janes84}                  & DDO                            & $0.056$ & $0.006$ \nl
\citet{burstein86}               & BH map\tablenotemark{f}        & $0.035$\tablenotemark{g} & $0.005$ \nl
\citet{nissen87}                 & $uvby$--$\beta$\tablenotemark{b} & $0.032$ & $0.006$ \nl
\citet{hobbs91}                  & $T_{\rm eff}$--$(B-V)_0$       & $0.065$ & $0.018$ \nl
\citet{montgomery93}             & $UBV$                          & $0.05$  & $0.01$ \nl
\citet{schlegel98}               & Dust map                       & $0.032$ & $0.005$\tablenotemark{h} \nl
Weighted mean                    & \nodata                        & $0.041$ & $0.004$ \nl
                                 & \nodata                        & \nodata & ($0.010$)\tablenotemark{c}\nl
\hline
\multicolumn{3}{c}{NGC~2516}\nl
\hline
\citet{eggen72}                  & $UBV$                          & $0.125$ & $0.025$ \nl
\citet{feinstein73}              & Sp-type, $UBV$                 & $0.116$ & $0.004$ \nl
\citet{snowden75}                & $uvby$--$\beta$\tablenotemark{b} & $0.120$ & $0.014$ \nl
\citet{mermilliod81a}            & $UBV$                          & $0.11 $ & $0.01$\tablenotemark{a} \nl
\citet{nicolet81}                & Geneva                         & $0.114$ & $0.012$\nl
\citet{eggen83}                  & $uvby$--$\beta$\tablenotemark{b} & $0.118$ & $0.004$\nl
\citet{verschoor83}              & $VBLUW$                        & $0.127$ & $0.005$\nl
\citet{nissen88}                 & $uvby$--$\beta$\tablenotemark{b} & $0.109$ & $0.007$\nl
\citet{dachs89}                  & Sp-type, $UBV$                 & $0.12 $ & $0.004$\nl
\citet{sung02}                   & $UBV$                          & $0.112$ & $0.006$\nl
Weighted mean                    & \nodata                        & $0.117$ & $0.002$\nl
                                 & \nodata                        & \nodata & ($0.005$)\tablenotemark{c}
\enddata
\tablenotetext{a}{Assigned value in this paper.}
\tablenotetext{b}{$E(B -  V) = E(b -  y) / 0.74$ \citep{cousins85}.}
\tablenotetext{c}{Standard deviation of the measurements.}
\tablenotetext{d}{Average over the cluster field.}
\tablenotetext{e}{Average over the HI hole (see text).}
\tablenotetext{f}{\citet{burstein82}.}
\tablenotetext{g}{Corrected for the systematic difference with \citet{schlegel98} of 0.02 mag.}
\tablenotetext{h}{Quoted precision (16\%).}
\end{deluxetable*}

Table~\ref{tab:ebv.lit} lists reddening estimates from a variety
of sources. For the Pleiades, we adopted a uniform reddening
inferred from the \ion{H}{1} hole measurement by \citet{breger86}.
Although small parts of the cluster show higher reddening
\citep[e.g.,][]{eggen50,mitchell57}, the stars in these regions
constitute only a small fraction of our Pleiades sample, and most of
these were tagged as outliers from our filtering algorithm
(\S~\ref{sec:filtering}). We compared magnitudes and colors derived from
individual $E(B - V)$ corrections for 157 stars
\citep{breger86,stauffer87,soderblom93} with those from the uniform
reddening, but the differences were negligible in our MS fitting range
(\S~\ref{sec:msfit}). The reddening measurements for the Pleiades
from other studies in Table~\ref{tab:ebv.lit} likely represent
a cluster average including the CO cloud region.

\section{CALIBRATED ISOCHRONES}

We used the YREC to construct stellar
evolutionary tracks at $-0.3 \leq {\rm [Fe/H]} \leq +0.2$ in 0.1 dex
increments. At each metal abundance, we ran a grid of masses from
0.2 to 8 $M_\odot$ in $0.05 - 1 M_\odot$ increments depending on
the stellar mass. The scaled solar abundance mix from \citet{grevesse98}
and the helium enrichment parameter ${\Delta Y} / {\Delta Z} = 1.2$ from
non-diffusion models were used; see \citet{sills00} and Paper~I for
detailed information. The tracks were interpolated to generate
theoretical isochrones at stellar ages from 20 Myr to 4 Gyr.

Stellar luminosities and $T_{\rm eff}$ were initially converted to
$V$, $B - V$, $V - I_C$, and $V - K_s$ from the relations by
\citet{lejeune97,lejeune98}. To obtain finer grids of the isochrones,
we increased the $M_V$ spacing of the isochrone points using a cubic
spline to match the $\Delta M_V = 0.05$ spacing of our tabulated
color-$T_{\rm eff}$ corrections in Paper~II. We then linearly interpolated
the isochrones in both {\rm [Fe/H]} and age at a fixed $M_V$. We finally
applied the empirical color corrections, as defined in Paper~II and
updated in the Appendix of this paper. We assumed that the correction
table is valid over the entire model grid in {\rm [Fe/H]} and age.

The isochrones constructed in this way are available online.
\footnote{See http://www.astronomy.ohio-state.edu/iso/.}

\section{DISTANCES AT FIXED REDDENING}

We now proceed to derive the cluster distances, adopting the reddening
values from Table~\ref{tab:ebv.lit}. We do this in two ways, either
holding the metallicity fixed at the spectroscopic values
(Table~\ref{tab:feh.lit}) or determining a photometric metallicity
that brings the distances from different CMDs into agreement. In either
case, we show that the calibrated isochrones improve the internal
precision of distance estimates from the three CMDs with $B - V$, $V - I_C$,
and $V - K_s$ as color indices, and $V$ as a luminosity index; hereafter
$(B - V, V)$, $(V - I_C, V)$, and $(V - K_s, V)$, respectively.

\subsection{Photometric Filtering}\label{sec:filtering}

Before fitting isochrones, we identified and removed stars that
are far from the MS in comparison to the sizes of their photometric
errors. Such stars could be either foreground/background objects
or cluster binaries that stand off from the MS.

The filtering process iteratively identifies the MS as the locus
of points of maximum density on the CMDs, independently of the
isochrones. As an initial step, we identified the MS by hand,
and then removed stars more than 1 mag away in $V$. This was necessary
in particular for the analysis in NGC~2516, which shows many faint
background stars in its CMDs (see below) that would complicate finding
the density maxima. The generous rejection criterion of 1 mag was
chosen so that all cluster binaries would be preserved at this stage.

At each step of the iteration, data points on each CMD were sorted
by $V$ magnitude into non-overlapping groups: each group contained
$\sqrt{N}$ points, where $N$ is the number of data points within
a color range used in the filtering.  The median colors in each group
were computed, and then the resulting run of points ($V$ vs. color)
was made into a smooth curve by averaging each point with the linear
interpolation of adjacent points.

With the above trial MS, we computed $\chi^2$ from each CMD as
\begin{equation}
 \chi^2 =
    \sum_{i=1}^N \chi_i^2 =
    \sum_{i=1}^N \frac {(\Delta V_i)^2 }
         {\sigma_{V,i}^2 + (\gamma_i \sigma_{c,i})^2 + \sigma_0^2},
\label{eq:chifilter}
\end{equation}
where $\Delta V_i$ is the $V$ difference between the $i^{th}$ data
point and MS at the color of the star and $\sigma_{V,i}$ and
$\sigma_{c,i}$ are the photometric errors in $V$ and color,
respectively. The error in the color contributes to the error in
$\Delta V_i$ by the slope of the curve, $\gamma_i$, which was
evaluated at the star's color. Because the above constructed curve
would not precisely trace the observed MS in the presence of
outliers, we added $\sigma_0$ in quadrature to the propagated
photometric errors in the denominator. We adjusted the value of
$\sigma_0$ so that the total $\chi^2$ is equal to $N$.
Initially, we rejected all data points as the MS outliers if the
$\chi_i^2$ (the individual contribution to $\chi^2$) is greater than
9 (corresponding to a 3 $\sigma$ outlier). We repeated adjusting
$\sigma_0$ and rejecting outliers with the reduced set of data
points until there remained no point with $\chi_i^2$ greater
than the above threshold value. We combined the results from all
three CMDs and rejected stars if they were tagged as an outlier
from any one of the CMDs.

We imposed an additional condition for the convergence of the algorithm
because cluster binaries and foreground/background stars are often
substantial populations near the MS. We compared the rms of
$\Delta V_i$ with its median of all absolute deviations (MAD).
The MAD of a quantity $x_i$ is defined by
\begin{equation}
 {\rm MAD} = 1.483\ {\rm median} (| x_i - {\rm median} (x_i) |),
 \label{eq:mad}
\end{equation}
where the correction factor 1.483 makes the estimator consistent
with the standard deviation for a normal distribution
\citep{rousseeuw90}. The MAD is a more robust estimator of
the dispersion than the standard deviation in the presence of
outliers. Therefore, we reduced the size of the $\chi_i^2$ threshold
value if the fractional difference between the rms and the MAD
of $\Delta V_i$ is larger than 5\%, assuming
${\rm median} (\Delta V_i) = 0$. We repeated the above filtering
steps by constructing a new MS curve from the remaining data set.

\begin{figure}
\epsscale{1.2}
\plotone{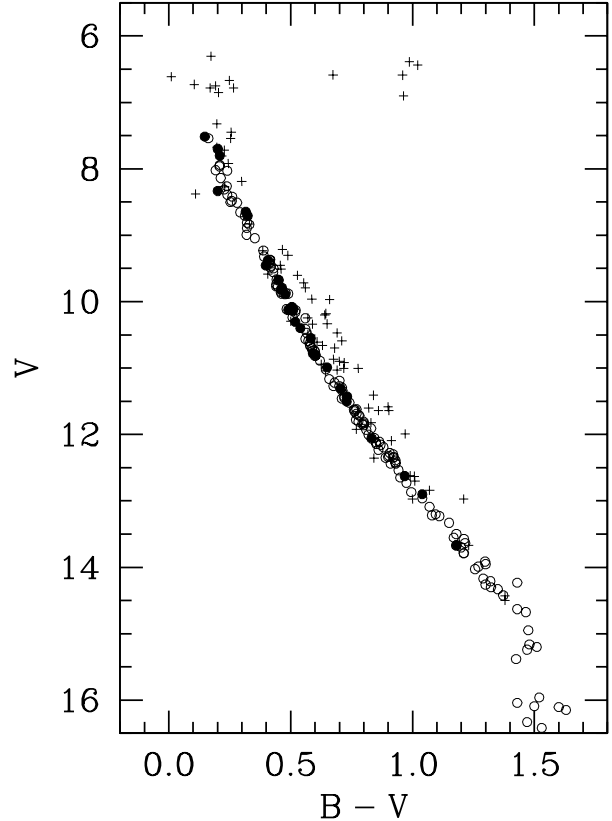}
\caption{Result of the filtering process in Praesepe. All known cluster
binaries were included to test the filtering algorithm. The plus signs
are rejected points from the filtering, and open circles are those
remaining. Filled circles are known binaries that remained after
the filtering. Stars brighter than $V = 7$ were excluded before
the filtering algorithm was applied. \label{fig:filter}}
\end{figure}

Figure~\ref{fig:filter} illustrates the result of the filtering
algorithm in Praesepe. Here we include all known cluster binaries but
exclude stars with a low membership probability (\S~\ref{sec:membin}).
We used all three CMDs in the filtering, but only the $(B - V, V)$
diagram is shown. The plus signs are rejected points from the filtering,
and open circles are those remaining. Some of the stars on the MS were
rejected because they were filtered in other CMDs. Although many
binaries remain after the filtering ({\it filled circles}), their proximity
to the MS would have a minor impact on the derived distance.
The bias in the distance due to the remaining binaries is discussed
in \S~\ref{sec:bin} from artificial cluster CMD tests.

\subsection{Isochrone Fitting and the Photometric Metallicity}
\label{sec:msfit}

The cluster distances were found by fitting isochrones over a range of
metallicity with the adopted age of each cluster (\S~\ref{sec:sample}).
The isochrones were reddened to the $E(B - V)$ values in
Table~\ref{tab:ebv.lit}. We adopted a reddening prescription for
the broadband colors by \citet{bessell98}, who have assumed
the extinction law of \citet{mathis90}. Their reddening formulae were
computed from ATLAS9 synthetic stellar photometry for a wide range of
$T_{\rm eff}$, and include color-dependent reddening resulting from
shifts in the effective wavelengths of broadband filters.
The Bessell et~al. formulae give reddening and extinction values for
$E(B - V) = 0.30$, so we linearly rescaled them according to the assumed
cluster reddening. The color transformation by
\citet{carpenter01}\footnote{Updated color transformations 
2MASS All-Sky Data Release can be found at\\
http://www.ipac.caltech.edu/2mass/releases/allsky/doc/sec6\_4b.html.}
was used to compute the reddening in $V - K_s$. For zero-color
stars, we found $R_V \equiv {{A_V} / {E(B - V)}} = 3.26$, $R_{VI}
\equiv E(V - I_C) / E(B - V) = 1.32$, and $R_{VK} \equiv E(V - K_s)
/ E(B - V) = 2.91$. For stars in the middle of our MS-fitting
range, $(B - V)_0 = 0.8$, we found $R_V = 3.44$, $R_{VI} =
1.37$, and $R_{VK} = 3.04$.

For a given isochrone, we computed each individual star's distance
modulus ($\mu_i$), and defined the cluster distance modulus in each CMD
as the ``unweighted'' median of $\mu_i$, i.e., $(m - M)_0 \equiv
{\rm median} (\mu_i)$. We computed the fitting error in the distance
modulus on each CMD as
\begin{equation}
 \sigma_{(m - M)} = \max
 \left[ \sigma_{\rm phot}, \frac{{\rm MAD} (\mu_i)}{\sqrt {N}} \right],
\end{equation}
where $N$ is the total number of data points used in the fit. Here
\begin{equation}
 {1 \over \sigma_{\rm phot}^2} = \sum^N_{i=1} {1 \over
 {\sigma^2_{V,i} + (\gamma_i \sigma_{c,i})^2} },
 \label{eq:sigphot}
\end{equation}
where the same notation is used as in equation~(\ref{eq:chifilter}) except
that $\gamma_i$ is the isochrone slope at the color of the star. We
further computed a ``weighted'' mean and a ``weighted'' median of
the distance modulus, and discuss the difference between the three
distance estimates along with other systematic errors in \S~\ref{sec:sys}.

For a given set of {\rm [M/H]}, age, and $E(B - V)$, we fit
isochrones in $(B - V, V)$, $(V - I_C, V)$, and $(V - K_s, V)$ over
$0.4 \leq (B - V)_0 \leq 1.3$. This color range is where the Hyades
calibration is most reliable (Paper~II). There were only a few Hyades
members blueward of the range, and the cool end was set by the magnitude
limit of the {\it Hipparcos} mission. The range corresponds to
$0.48 \leq (V - I_C)_0 \leq 1.48$ and $0.98 \leq (V - K_s)_0 \leq 3.16$.
For non-zero reddening, the fitting ranges were made correspondingly redder.

\begin{figure}
\epsscale{1.2}
\plotone{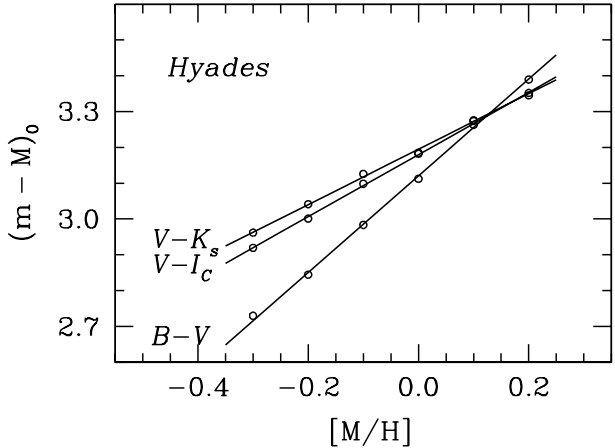}
\caption{Example of determining a photometric metallicity in the Hyades.
Open circles are the distances derived from CMDs with $B - V$,
$V - I_C$, and $V - K_s$ as a function of isochrone metallicity.
The solid lines connecting these points are least-squares fits, and
the labels to the left of the lines indicate the corresponding color
index. An age of 550 Myr and $E(B - V) = 0.000$ were assumed.
The photometric metallicity is defined as the average {\rm [M/H]} where
the $B - V$ line crosses the $V - I_C$ and $V - K_s$ lines.
\label{fig:hy.md}}
\end{figure}

We define the photometric metallicity ${\rm [M/H]}_E$ as the one
that brings distances from three CMDs into statistical agreement
\citep[e.g.,][]{pinsono98,stello01,terndrup02}.
Figure~\ref{fig:hy.md} shows how this process works for the
Hyades, our calibrating cluster. The open circles display the
derived distances as a function of isochrone metallicity in each
of the three CMDs with $E(B - V) = 0.000$ at an age of 550 Myr.
The lines connecting these points are least-squares fits,
and the labels to the left of the lines indicate the corresponding
color index. The slope in the $(B - V, V)$ is larger than in
the other two CMDs, indicating a greater sensitivity of
the isochrone luminosity to the metal abundance. We
define ${\rm [M/H]}_E$ as the weighted average of the two
metallicities at which the $(B - V, V)$ distance agrees with that
from $(V - I_C, V)$ and from $(V - K_s, V)$, respectively. For the
Hyades in Figure~\ref{fig:hy.md}, we derive the photometric metallicity
of ${\rm [M/H]}_E = +0.13 \pm 0.02$, which is naturally the same as
the originally assumed value in the isochrone calibration.

With this definition of the photometric metallicity, we can derive two
distances at the adopted reddening. The first of these, designated as
$(m - M)_{0,S}$, is the weighted average distance modulus from the three
CMDs at the spectroscopic {\rm [Fe/H]}. The second distance,
$(m - M)_{0,E}$, is the value determined at the photometric metallicity,
${\rm [M/H]}_E$.

\subsection{Results}\label{sec:results}

\begin{figure*}
\epsscale{0.95}
\plotone{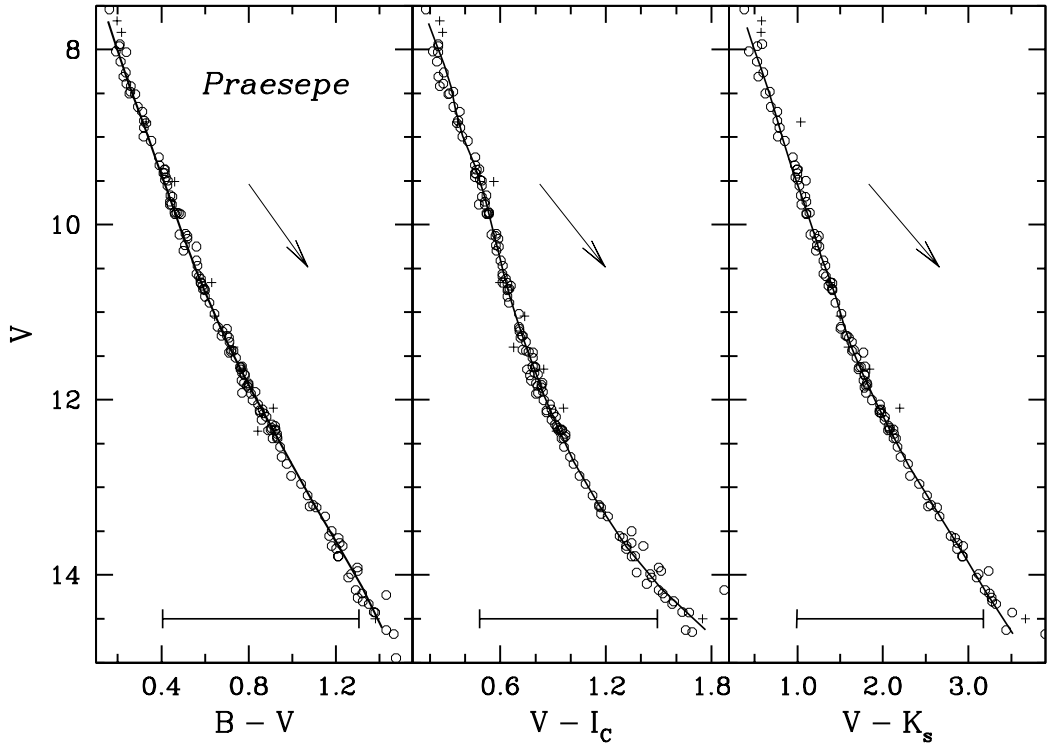}
\caption{CMDs of Praesepe, after excluding known binaries and stars with
low membership probability. Plus signs are photometrically rejected data
points from the filtering algorithm, and open circles are those remaining.
The solid lines are empirically calibrated isochrones with spectroscopic
metallicity (Table~\ref{tab:feh.lit}), which were adjusted for
the reddening with the literature value (Table~\ref{tab:ebv.lit}).
Fitting ranges are shown as horizontal bars. The arrow denotes
the direction of reddening vectors.\label{fig:pr.cmd}}
\end{figure*}

\begin{figure*}
\epsscale{0.95}
\plotone{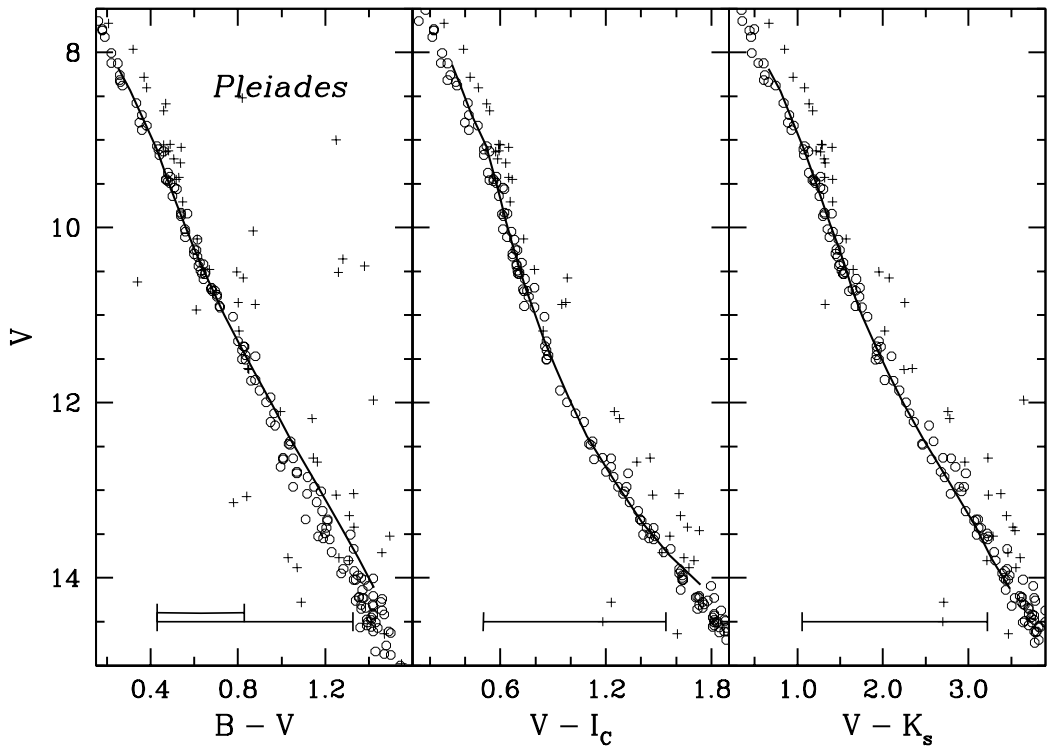}
\caption{Same as Fig.~\ref{fig:pr.cmd}, but
for the Pleiades. As discussed by \citet{stauffer03}, the Pleiades
K dwarfs are bluer than the given isochrone in $(B - V, V)$, but
not in the other two CMDs (see text). The distance modulus in
$(B - V, V)$ was derived at $0.4\leq (B - V)_0 \leq 0.8$ as
shown by the shorter horizontal bar. \label{fig:pl.cmd}}
\end{figure*}

\begin{figure*}
\epsscale{0.95}
\plotone{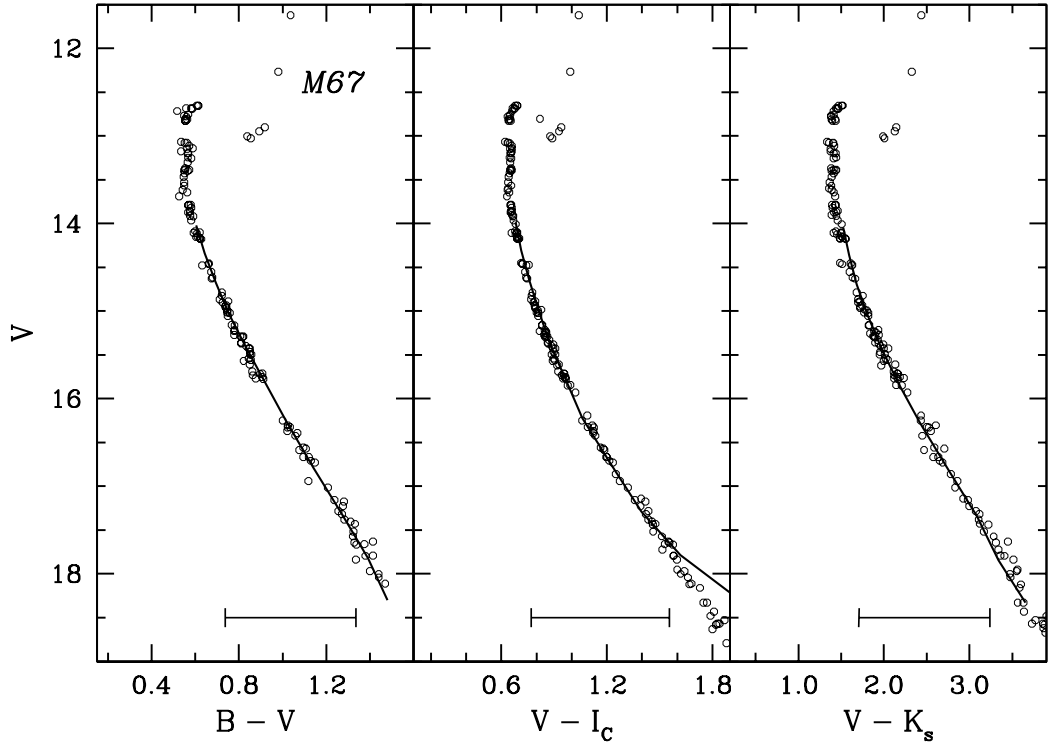}
\caption{Same as Fig.~\ref{fig:pr.cmd}, but for M67 with photometry
from S04. Shown are those stars designated as single cluster members by
S04. The fit was performed over $0.7\leq (B - V)_0 \leq1.3$ and
corresponding color ranges in the other two CMDs to avoid the steeply
rising part of the upper MS.\label{fig:67.cmd}}
\end{figure*}

\begin{figure*}
\epsscale{0.95}
\plotone{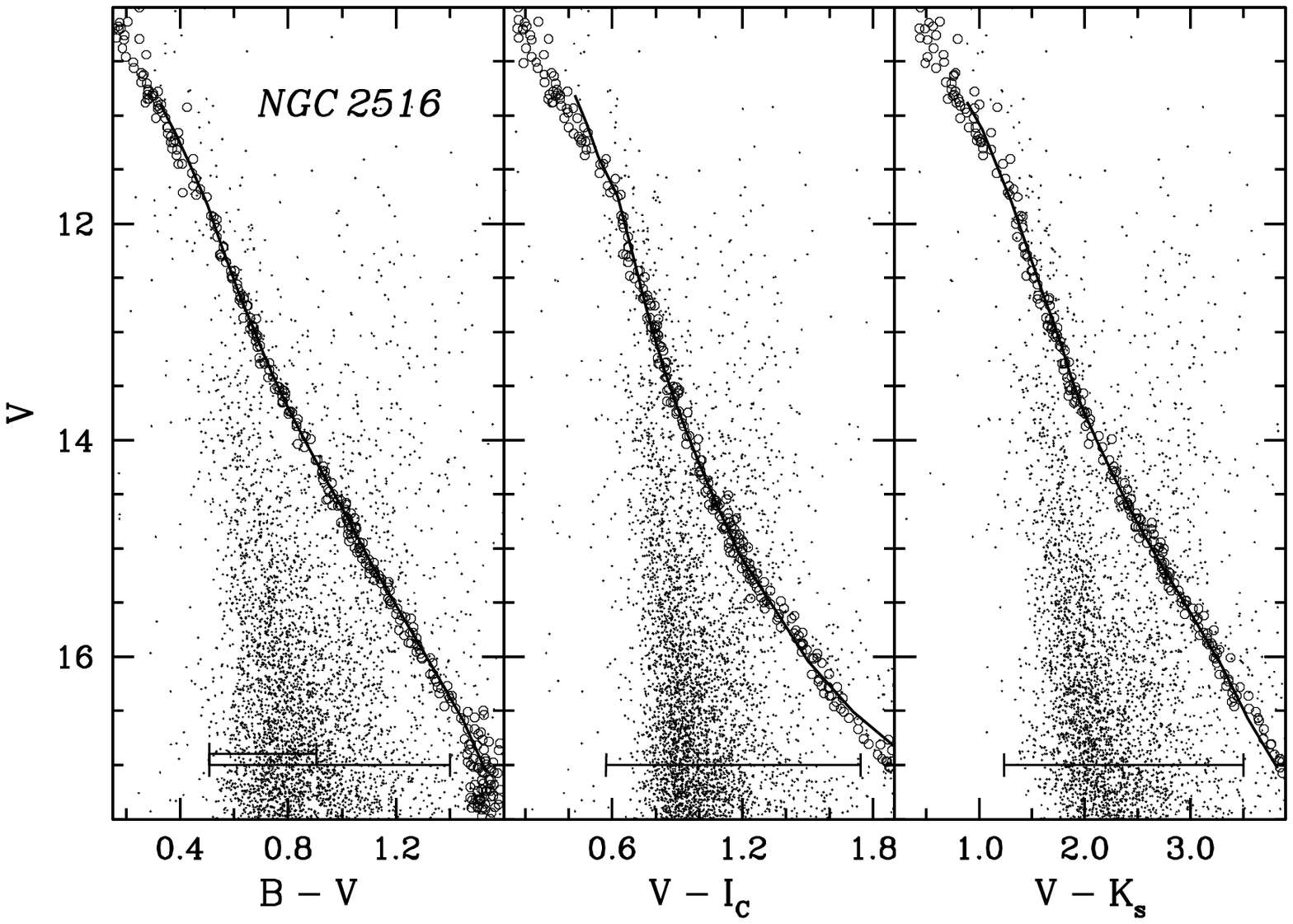}
\caption{ Same as Fig.~\ref{fig:pr.cmd}, but for NGC~2516 with
photometry from JTH01. The fitting range in $(B - V, V)$ was
truncated in the red part as in the Pleiades CMD (see text).
\label{fig:25.cmd}}
\end{figure*}

Figures~\ref{fig:pr.cmd}-\ref{fig:25.cmd} display the CMDs for
each cluster. For Praesepe and the Pleiades, known binaries and
nonmembers were excluded before we applied the photometric
filtering. Stars that remained after the filtering are shown as
open points, and the rejected stars are shown as plus signs. With the
extensive binary and membership information, the CMDs of Praesepe
(Fig.~\ref{fig:pr.cmd}) are dominated by single cluster members, and
there are few stars rejected by the photometric filtering.
The CMDs of the Pleiades (Fig.~\ref{fig:pl.cmd}) show that the
filtering routine identified likely cluster binaries effectively.
The single cluster members from S04 are shown in the CMDs of M67
(Fig.~\ref{fig:67.cmd}), none of which were rejected by the
filtering algorithm. The CMDs of NGC~2516 (Fig.~\ref{fig:25.cmd})
are from JTH01 and have a large number of foreground/background stars
as expected in an area survey (for clarity, the point size for
the rejected stars was reduced on the CMDs as compared to the other plots).

Isochrones at the spectroscopic metallicities are overlaid as
solid lines in each figure, and in most cases they are in excellent
agreement with the observed shapes of the MS. In particular, the match
to the Praesepe MS (Fig.~\ref{fig:pr.cmd}), which has an age and
metallicity nearly identical to the Hyades, is good over almost
7 mag in $V$. We interpret this agreement as indicating
that the Hyades-based corrections to the color-$T_{\rm eff}$ relations
defined in Paper~II apply to all clusters, or at least to those
with metallicities not too different from that of the Hyades.

There are a few cases in which the match to the MS shape is not as
good as it is in Praesepe. In the Pleiades (Fig.~\ref{fig:pl.cmd}),
stars with $B - V \ga 0.9$ are considerably bluer than the isochrone
in $(B - V, V)$, although not in the other two CMDs. This ``blue K dwarf''
phenomenon was discussed by \citet{jones72}, \citet{landolt79}, and
\citet{vanleeuwen87}, and at greater length by \citet{stauffer03},
who attributed it to stellar temperature inhomogeneities caused by
cool spots and plage areas in rapidly rotating young stars. From
the JTH01 photometry, it would appear that NGC~2516, which is about
40\% older than the Pleiades, does not share this phenomenon with
the Pleiades (see Fig.~\ref{fig:25.cmd}). However, we show in
\S~\ref{sec:activity} that the usage of the S02 data comes to
a different conclusion with supporting evidence from stellar rotational
velocities. The data in $(V - I_C, V)$ for all clusters are somewhat
bluer than the isochrone redward of $V - I_C \approx 1.6$; this
indicates a possible systematic error in the Hyades calibration for
these red stars, probably resulting from a paucity of Hyades stars with
accurately measured parallaxes in this color range. In M67
(Fig.~\ref{fig:67.cmd}), there may be a larger deviation because
the color calibration employed by S04 did not extend as far to
the red as this.

The distance was independently derived from each CMD over the color
range shown as a horizontal bar in each panel. The blue end of
the color ranges for M67 was increased to $(B - V)_0 = 0.7$ to avoid
the steeply rising parts of the MS, as the resulting distance would
be sensitive to the choice of cluster age. For the Pleiades and
NGC~2516, the red end of the color range in $(B - V, V)$ was decreased
to $(B - V)_0 = 0.8$ to avoid systematic errors from the blue K dwarf
phenomenon.

\begin{deluxetable*}{lcccccc}
\tablewidth{0pt}
\tablecaption{MS Fitting Distance and Photometric Metallicity at Literature $E(B -  V)$
\label{tab:dist}}
\tablehead{
  \colhead{} &
  \multicolumn{3}{c}{$(m -  M)_0$} &
  \colhead{} &
  \colhead{} &
  \colhead{} \nl
  \cline{2-4}
  \colhead{Data / Constraint} &
  \colhead{$(B -  V, V)$} &
  \colhead{$(V -  I_C, V)$} &
  \colhead{$(V -  K_s, V)$} &
  \colhead{$(m -  M)_{0,S}$} &
  \colhead{${\rm [M/H]}_E$} &
  \colhead{$(m -  M)_{0,E}$}
}
\startdata
\multicolumn{7}{c}{Praesepe}\nl
\hline
known binaries excluded &
 $6.267\pm0.008$ & $6.291\pm0.008$ & $6.336\pm0.007$ & $6.303\pm0.021$ & $+0.207\pm0.030$ & $6.395\pm0.030$ \nl
known binaries included &
 $6.262\pm0.008$ & $6.279\pm0.008$ & $6.327\pm0.007$ & $6.293\pm0.020$ & $+0.186\pm0.030$ & $6.367\pm0.031$ \nl
\hline
\multicolumn{7}{c}{Pleiades}\nl
\hline
known binaries excluded &
 $5.658\pm0.010$ & $5.623\pm0.013$ & $5.669\pm0.017$ & $5.649\pm0.013$ & $+0.021\pm0.032$ & $5.627\pm0.039$ \nl
known binaries included &
 $5.660\pm0.009$ & $5.639\pm0.013$ & $5.677\pm0.016$ & $5.657\pm0.009$ & $+0.035\pm0.028$ & $5.651\pm0.036$ \nl
\hline
\multicolumn{7}{c}{M67}\nl
\hline
S04                     &
 $9.631\pm0.008$ & $9.601\pm0.007$ & $9.654\pm0.011$ & $9.622\pm0.015$ & $-0.002\pm0.062$ & $9.620\pm0.054$ \nl
MMJ93                   &
 $9.630\pm0.010$ & $9.505\pm0.009$ & $9.626\pm0.008$ & $9.584\pm0.042$ & $-0.032\pm0.183$ & $9.562\pm0.163$ \nl
MMJ93 / S04 stars\tablenotemark{a} &
 $9.612\pm0.013$ & $9.529\pm0.010$ & $9.631\pm0.011$ & $9.584\pm0.033$ & $+0.017\pm0.138$ & $9.599\pm0.109$ \nl
\hline
\multicolumn{7}{c}{NGC~2516}\nl
\hline
JTH01                   &
 $8.125\pm0.007$ & $8.040\pm0.009$ & $8.100\pm0.007$ & $8.096\pm0.023$ & $-0.052\pm0.053$ & $8.034\pm0.058$ \nl
JTH01 / RV + X-ray\tablenotemark{b} &
 $8.115\pm0.010$ & $8.041\pm0.014$ & $8.077\pm0.021$ & $8.088\pm0.023$ & $-0.074\pm0.030$ & $7.995\pm0.037$ \nl
S02                  &
 $8.098\pm0.014$ & $8.035\pm0.013$ & $8.064\pm0.012$ & $8.065\pm0.017$ & $-0.042\pm0.024$ & $8.015\pm0.026$ \nl
S02 / RV + X-ray\tablenotemark{b} &
 $8.081\pm0.013$ & $7.960\pm0.022$ & $8.063\pm0.022$ & $8.052\pm0.034$ & $-0.059\pm0.081$ & $7.979\pm0.092$
\enddata
\tablecomments{All subsamples of data were filtered independently before fitting isochrones.}
\tablenotetext{a}{Single cluster members from S04 with MMJ93 photometry.}
\tablenotetext{b}{RV members from \citet{terndrup02}, and X-ray detected sources from \citet{damiani03}.}
\end{deluxetable*}

In Table~\ref{tab:dist} we assemble our derived cluster distances
and the photometric metallicities. The first column lists the source
of photometry, including particular selections of subsamples of the data.
We applied the photometric filtering to each subsample before fitting
isochrones. The second through fourth columns display the distance modulus
found from each CMD, where the spectroscopic metallicity and reddening were
taken from Tables~\ref{tab:feh.lit} and \ref{tab:ebv.lit}, respectively.
The weighted average distance, $(m - M)_{0,S}$, is shown in the fifth
column. The last two columns display the photometric metallicity,
${\rm [M/H]}_E$, and the distance modulus, $(m - M)_{0,E}$, at
this metallicity. The error in $(m - M)_{0,S}$ is either
the propagated one from the errors of the individual distances
or the s.e.m.\ of these distance estimates, whichever is larger.
The error in $(m - M)_{0,E}$ additionally includes the propagated error
from ${\rm [M/H]}_E$.

\begin{figure}
\epsscale{1.2}
\plotone{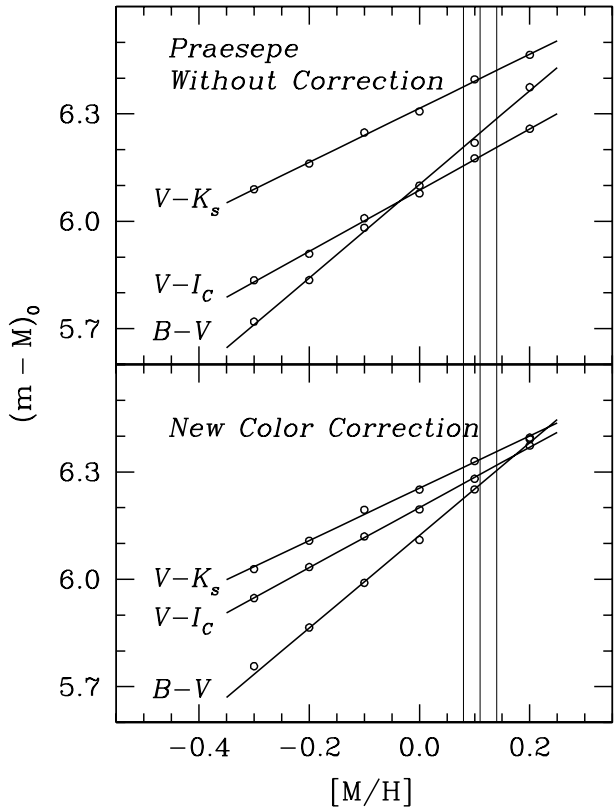}
\caption{Distance vs. isochrone metallicity for Praesepe with
the \citet{lejeune97,lejeune98} color-$T_{\rm eff}$ relation ({\it top})
and with our new empirical correction ({\it bottom}). The reddening
value in the literature (Table~\ref{tab:ebv.lit}) was assumed. Circles
and lines have the same meaning as in Fig.~\ref{fig:hy.md}. Vertical
lines show the spectroscopic metallicity and its 1 $\sigma$ error
(Table~\ref{tab:feh.lit}).\label{fig:pr.md}}
\end{figure}

The standard errors of the average distance modulus at the spectroscopic
metallicities are typically on the order of 0.02 mag
(i.e., the individual distances in the three CMDs are consistent
to $\sim2\%$). In Figure~\ref{fig:pr.md} we show that this consistency
is primarily the result of the empirical color-$T_{\rm eff}$ corrections
derived in Paper~II. The top panel shows
the derived distances to Praesepe as a function of the metallicity
for isochrones that do not incorporate these corrections, while
the bottom panel shows what happens when the corrected isochrones are
employed. The vertical lines show the metallicity estimate for Praesepe
with its 1 $\sigma$ error (\S~\ref{sec:spec}). The internal precision
of the distance estimation with isochrones employing the Hyades empirical
calibration is about a factor of 5 better than with the uncorrected
isochrones.

The distances and photometric metallicities in Table~\ref{tab:dist}
are generally not sensitive to the selection of subsamples in each
cluster. For example in Praesepe and the Pleiades, leaving known binaries
in the sample before filtering changed the value of $(m - M)_{0,S}$ by only
about 0.01 mag, confirming that the filtering algorithm removed most of
the binaries that are brighter than the MS (e.g., Fig.~\ref{fig:filter}).
In M67, we compared the distances from the MMJ93 data, but also using
their photometry for only those stars identified as single members by
S04; we found a negligible difference. In NGC 2516, we computed
the distances from the full JTH01 or S02 catalogs, or selecting
only those stars identified as radial velocity (RV)
members \citep{terndrup02} or X-ray detected sources \citep{damiani03};
again making that selection reduced the distance modulus by only about
0.01 mag.

\begin{figure}
\epsscale{1.2}
\plotone{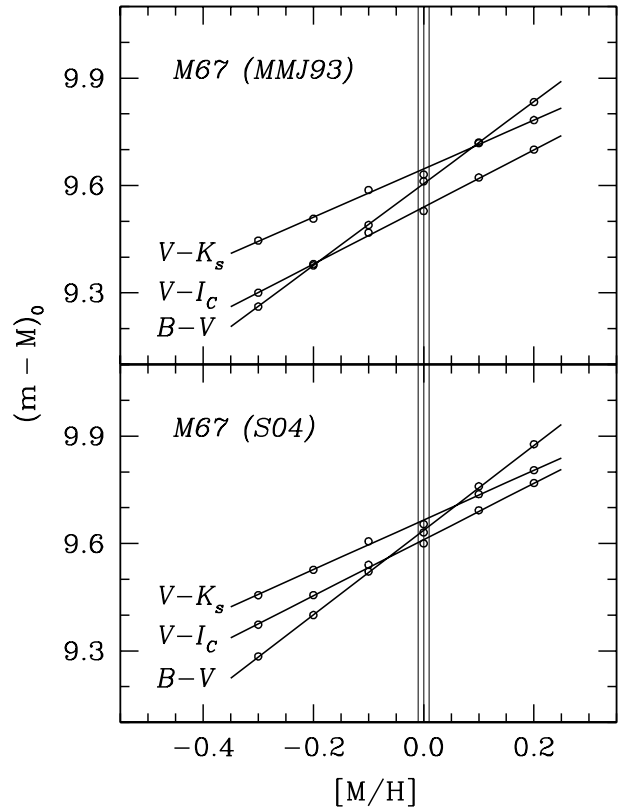}
\caption{Same as Fig.~\ref{fig:pr.md}, but for M67 with MMJ93
({\it top}) and S04 ({\it bottom}) photometry. The stars designated
as single cluster members by S04 were used in both cases.\label{fig:67.md}}
\end{figure}

\begin{figure}
\epsscale{1.2}
\plotone{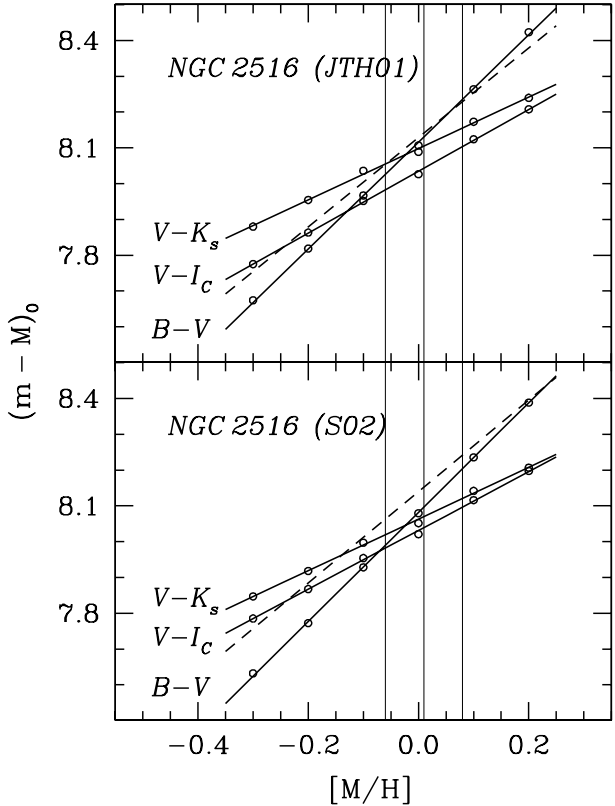}
\caption{Same as Fig.~\ref{fig:pr.md}, but for NGC~2516 with JTH01
({\it top}) and S02 ({\it bottom}) photometry. The fitting range
in $(B - V, V)$ was restricted to $0.4\le (B - V)_0 \le0.8$
as in the case of the Pleiades. The dashed line shows the result
when the full fitting range was used in this color.\label{fig:25.md}}
\end{figure}

\begin{figure}
\epsscale{1.2}
\plotone{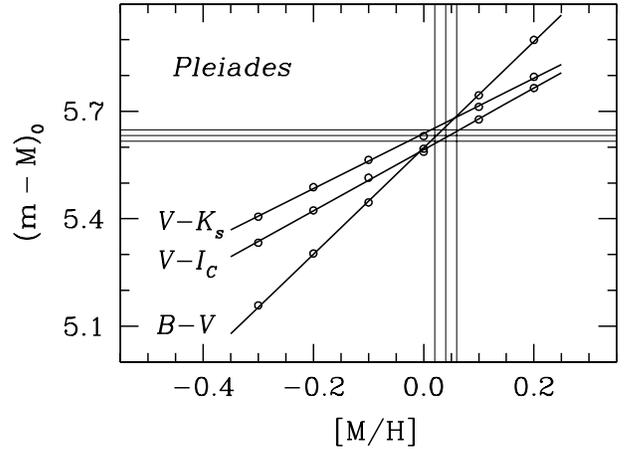}
\caption{Same as Fig.~\ref{fig:pr.md}, but for the Pleiades. Horizontal
lines represent the average geometric distance and its 1 $\sigma$ error
from Table~\ref{tab:hipp}.\label{fig:pl.md}}
\end{figure}

Figures~\ref{fig:67.md}--\ref{fig:pl.md} display the derived
distances as a function of metallicity for M67, NGC~2516, and
the Pleiades, respectively. The vertical lines on each plot show
the average {\rm [Fe/H]} and its 1 $\sigma$ error from
Table~\ref{tab:feh.lit}.
The distances for M67 (Fig.~\ref{fig:67.md}) are significantly less
consistent with the MMJ93 photometry than from the S04 data. This is
also seen in the larger errors in ${\rm [M/H]}_E$ and $(m - M)_{0,E}$
in Table~\ref{tab:dist}. This is mainly because the $V - I_C$ colors
in MMJ93 are (on average) 0.022 mag redder than in S04, which will
reduce the distance in the $(V - I_C, V)$ by about 0.07 mag since
the average slope of the MS in this color is about 3.

In Figure~\ref{fig:25.md} we explore the effects of the blue K dwarfs
in NGC~2516 by comparing the two photometric data sets. The distances in
$(B - V, V)$ derived over the truncated color range are shown as
a solid line, while the dashed line displays the results for fits
over $0.4 \leq (B - V)_0 \leq 1.3$. The fits in the other two CMDs
were done over the full color ranges. When the truncated color range
is used, both JTH01 and S02 data give distances and
photometric metallicities that are mutually consistent. However,
fitting on the S02 data over the full range of colors in $B - V$
reduces the photometric metallicity by $\sim0.12$ dex and
the average distance at this metallicity by $\sim0.09$ mag, while
the results from the JTH01 data remain consistent within 1 $\sigma$
independently of the fitting range. We discuss the blue K dwarfs
phenomenon in NGC~2516 at greater length in \S~\ref{sec:activity}.

In Figure~\ref{fig:pl.md} we display the fits for the Pleiades, and
also plot the average geometric distance and its 1 $\sigma$ error
from Table~\ref{tab:hipp} as the horizontal lines. The agreement
between the MS-fitting and geometric distances is superb. Our distance
for Praesepe at the spectroscopic metallicity also agrees with
existing geometric distance determinations: \citet{gatewood94} found
$(m - M)_0 = 6.42 \pm 0.33$ from ground-based parallaxes, and
\citet{loktin00} obtained $6.16 \pm 0.19$ from the moving cluster method.
In addition, there are two determinations of the distance to Praesepe
from {\it Hipparcos}: $(m - M)_0 = 6.37 \pm 0.15$ \citep{vanleeuwen99} and
$6.28^{+0.13}_{-0.12}$ \citep{robichon99}.  We compare our derived
distances with the {\it Hipparcos} measurements in \S~\ref{sec:hipp}.

\section{SIMULTANEOUS DETERMINATION OF CLUSTER PARAMETERS}\label{sec:chisqr}

In the previous section, we showed that the cluster metallicities
can be obtained from the MS fitting because the various color
indices have different sensitivities to the metallicity. However,
we can further constrain the reddening because the slope of the MS
becomes shallower below $M_V \approx 5.5$, especially in $(V - I_C, V)$
or $(V - K_s, V)$. Furthermore, the reddening vectors
are not parallel to the MS as shown in Figure~\ref{fig:pr.cmd},
so the derived distances depend on the reddening with different
degrees of sensitivity in the three CMDs.

To determine the metallicity, reddening, and distance simultaneously,
we minimized
\begin{equation}
 \chi^2_{\rm tot} = \sum_{j=1}^3 \sum_{i=1}^{N_j}
    \frac { ( \mu_{ij} - \bar{\mu} )^2 }
    { \left[ \sigma_{V,ij}^2 + (\gamma_{ij} \sigma_{c,ij})^2 \right]
    \times f_j^2 },
 \label{eq:chifit}
\end{equation}
where the same notation is used as in equation~(\ref{eq:chifilter}) except
that the subscript $j$ is used to indicate each of the three CMD.
The $N_j$ is the number of data points used in the fit on the $j^{th}$
CMD, and $\mu_{ij}$ are the individual distance moduli of the $i^{th}$
point. The $\bar{\mu}$ is the weighted average distance modulus from
the three CMDs.

The quantity $f_j$ in the denominator of equation~(\ref{eq:chifit})
is a renormalization factor to take into account underestimation of
the photometric errors. The value of $f_j$ was determined by comparing
the residuals about the best-fitting isochrone at the minimum
$\chi^2_{\rm tot}$ (see below) to the propagated distance error from
the photometric errors (eq.~[\ref{eq:sigphot}], $\sigma_{\rm phot}$).
This yields
\begin{equation}
  f_j = \max \left[ \frac{{\rm MAD} (\mu_{ij})}{\sqrt {N_j}}
  \frac{1}{\sigma_{{\rm phot},j}}, 1 \right].
\end{equation}
This renormalization was required in particular for the S04 data,
which had many more repeat measurements in $V$ and $I_C$ than in $B$,
so the quoted errors are much smaller in $V - I_C$ than in the other
colors. For these data, we found $f \approx 6$ in $(V - I_C, V)$,
indicating that the errors are greatly underestimated at least when
compared to the scatter about the best-fitting isochrone. Without
the renormalization factor, the majority of the weight in
the $\chi^2_{\rm tot}$ would be given to the $(V - I_C, V)$ CMD,
and the process would almost entirely ignore the information on
the other CMDs.

We searched for the minimum $\chi^2_{\rm tot}$ in the plane of
{\rm [M/H]} versus $E(B - V)$ using a downhill simplex method
\citep{press92}. The average distance modulus in equation~(\ref{eq:chifit})
was determined at each {\rm [M/H]} and $E(B - V)$.
Because the color ranges used in the fitting depend on the adopted
$E(B - V)$, we solved for the minimum $\chi^2_{\rm tot}$
at the $E(B - V)$ derived from the previous iteration to keep the same
number of data points for each iteration. Since we defined
$\chi^2_{\rm tot}$ with respect to the average distance modulus from
the three CMDs, the minimum $\chi^2_{\rm tot}$ yields a set of {\rm [M/H]}
and $E(B - V)$ that best describes all of the CMDs simultaneously.

\begin{figure*}
\epsscale{1.0}
\plotone{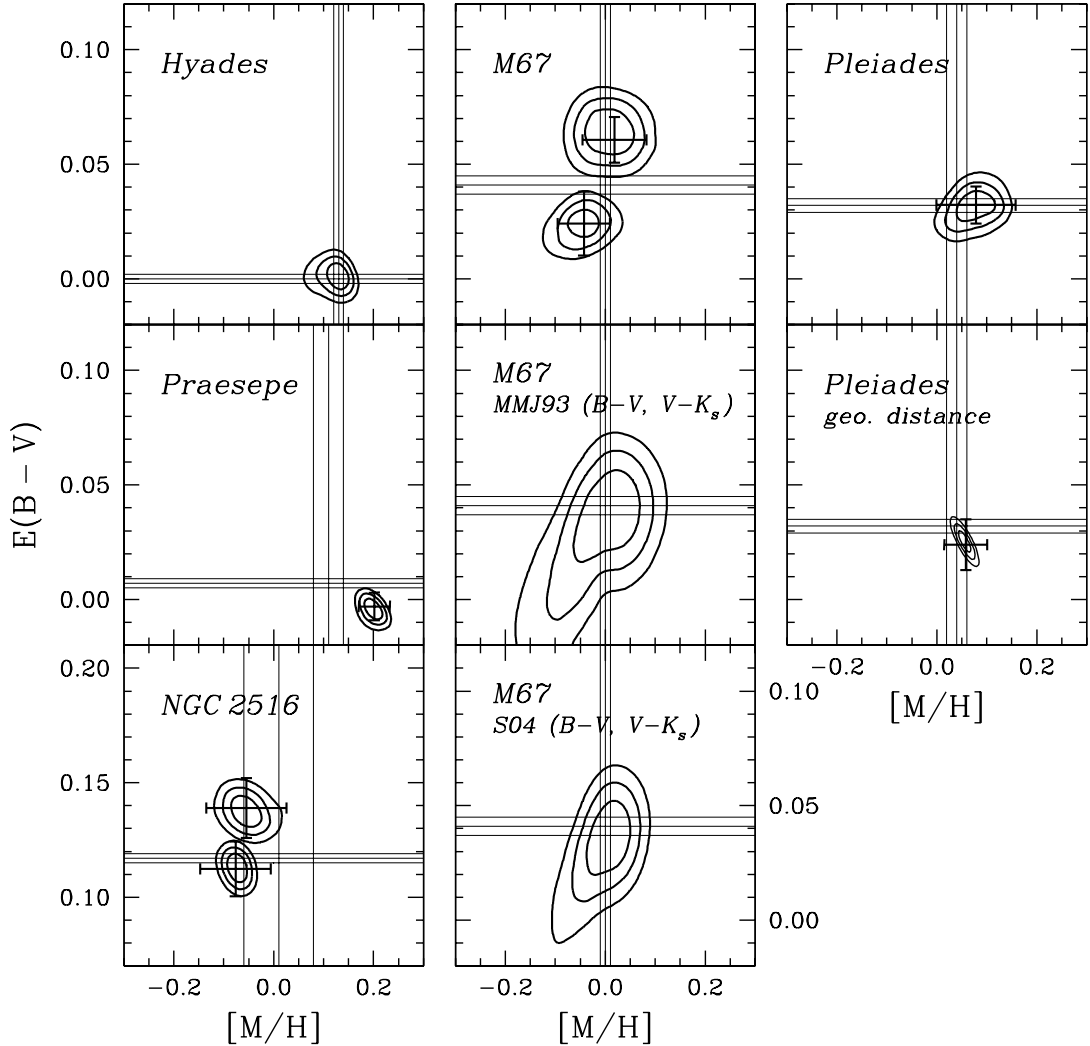}
\caption{Likelihood contours in {\rm [M/H]} vs.\ $E(B - V)$ shown
at $\Delta \chi^2_{\rm tot} =$ 3.53, 8.02, and 14.2 (1, 2, and 3 $\sigma$ for
3 degrees of freedom) relative to the minimum value of $\chi^2_{\rm tot}$
(eq.~[\ref{eq:chifit}]). Those for 2 degrees of freedom are shown for
the Pleiades at the geometric distance ({\it middle right}).
For NGC~2516 ({\it bottom left}), contours with the higher $E(B - V)$
are from the S02, and the others are from the JTH01 photometry.
For M67 ({\it top middle}), contours with the higher $E(B - V)$ are
from the MMJ93, and the others are from the S04 photometry.
Also shown are contours for the MMJ93 ({\it middle}) and the S04
({\it bottom middle}) in the case where only $(B - V, V)$ and
$(V - K_s, V)$ are used. Error bars represent 1 $\sigma$
systematic errors listed in Tables~\ref{tab:pr}-\ref{tab:25} excluding
fitting errors. The spectroscopic metallicities and reddening values
in the literature with 1 $\sigma$ errors (Tables~\ref{tab:feh.lit},
\ref{tab:ebv.lit}) are shown with vertical and horizontal lines,
respectively.\label{fig:me}}
\end{figure*}

Figure~\ref{fig:me} shows likelihood contours in the {\rm [M/H]}
versus $E(B - V)$ plane for all clusters in this paper and for the Hyades.
The $\chi^2$ surfaces were smoothed with a Gaussian kernel. Contours are
shown at $\Delta \chi^2_{\rm tot} =$ 3.53, 8.02, and 14.2 relative to
the minimum value of $\chi^2_{\rm tot}$, which correspond to 68.3\%,
95.4\%, and 99.7\% confidence levels (1, 2, 3 $\sigma$) for the 3
degrees of freedom, and the errors in both quantities were
assumed to be normally distributed without correlation.
The vertical lines on each panel display the spectroscopic metallicity
and its 1 $\sigma$ error, while the horizontal lines show the average
reddening in the literature with its 1 $\sigma$ error. The error bars in
each panel show the size of systematic errors, which is discussed
in the next section.

The solution for Praesepe (Fig.~\ref{fig:me}, {\it middle left panel}) yields
a value for the reddening that is slightly negative and a metallicity
that is $\sim 2\ \sigma$ higher than found spectroscopically. The negative
reddening may be a consequence of zero-point errors in the photometry
or else from an error in the Hyades reddening, which we assumed in
Paper~II as identically zero. The study of \citet{taylor80}, for
example, indicates $E(B - V) = +0.003 \pm 0.002$ for the Hyades. So,
if the Praesepe reddening was slightly less than the non-zero Hyades
reddening, the MS-fitting method would yield a negative value.

In Figure~\ref{fig:me} two solutions are shown for M67 ({\it top middle panel}):
the contours with the higher $E(B - V)$ are from the MMJ93 data, and
the others are from the S04. The large difference in $E(B - V)$ obtained
from these two samples is primarily due to the difference in
the $V - I_C$ values (Table~\ref{tab:phot.diff}). We illustrate this
further by showing the $\chi^2$ contours that result when only
the $B - V$ and $V - K_s$ data are included. The middle panel
shows the likelihood contours for the MMJ93 data, while the bottom
middle panel is for the S04. Here the two studies yield the same values
of metallicity and reddening, although with larger errors. We also display
two solutions for NGC~2516 in the bottom left panel of
Figure~\ref{fig:me}. The contours with the higher reddening are from
the SBLL03, and the others are from JTH01 photometry.

The top right panel of Figure~\ref{fig:me} shows the solution for
the Pleiades, while the panel below that shows the solution for
the metallicity and reddening that result if the average geometric
distance from Table~\ref{tab:hipp} is taken as a prior. Contours in
this panel are 1, 2, and 3 $\sigma$ confidence levels for the 2
degrees of freedom. This demonstrates that good photometry and
parallaxes together can provide strong constraints on the metal
abundance of a cluster.

\begin{deluxetable*}{lccc}
\tablewidth{0pt}
\tablecaption{$\chi^2$ Solutions\label{tab:chi}}
\tablehead{
  \colhead{Data / Constraint} &
  \colhead{$(m -  M)_{0,{\chi^2}}$} &
  \colhead{${\rm [M/H]}_{\chi^2}$} &
  \colhead{$E(B -  V)_{\chi^2}$}
}
\startdata
\multicolumn{4}{c}{Praesepe}\nl
\hline
known binaries excluded             & $6.365\pm0.025$ & $+0.202\pm0.020$ & $-0.003\pm0.005$\nl
known binaries included             & $6.357\pm0.026$ & $+0.192\pm0.020$ &  $0.002\pm0.004$\nl
\hline
\multicolumn{4}{c}{Pleiades}\nl
\hline
known binaries excluded               & $5.693\pm0.053$ & $+0.079\pm0.040$ & $0.032\pm0.007$\nl
At geometric distance\tablenotemark{a} & $5.632\pm0.017$\tablenotemark{b}
                                                        & $+0.058\pm0.012$ & $0.024\pm0.004$\nl
known binaries included               & $5.667\pm0.054$ & $+0.059\pm0.040$ & $0.027\pm0.007$\nl
At geometric distance\tablenotemark{c} & $5.632\pm0.017$\tablenotemark{b}
                                                        & $+0.049\pm0.012$ & $0.025\pm0.004$\nl
\hline
\multicolumn{4}{c}{M67}\nl
\hline
S04                                 & $9.558\pm0.034$ & $-0.043\pm0.032$ & $0.024\pm0.006$\nl
MMJ93                               & $9.603\pm0.051$ & $-0.061\pm0.035$ & $0.078\pm0.006$\nl
MMJ93 / S04 stars\tablenotemark{d}  & $9.638\pm0.053$ & $+0.019\pm0.050$ & $0.061\pm0.010$\nl
\hline
\multicolumn{4}{c}{NGC~2516}\nl
\hline
JTH01                               & $8.001\pm0.038$ & $-0.076\pm0.020$ & $0.112\pm0.007$\nl
JTH01 / RV + X-ray\tablenotemark{e} & $7.939\pm0.059$ & $-0.133\pm0.040$ & $0.120\pm0.010$\nl
S02                              & $8.064\pm0.038$ & $-0.055\pm0.030$ & $0.139\pm0.007$\nl
S02 / RV + X-ray\tablenotemark{e}& $8.038\pm0.091$ & $-0.050\pm0.060$ & $0.134\pm0.015$
\enddata
\tablecomments{All subsamples of data were filtered independently before fitting isochrones.}
\tablenotetext{a}{Solution at the average geometric distance after excluding known binaries.}
\tablenotetext{b}{Average geometric distance from Table~\ref{tab:hipp}.}
\tablenotetext{c}{Solution at the average geometric distance with known binaries included in the data.}
\tablenotetext{d}{Single cluster members from S04 with MMJ93 photometry.}
\tablenotetext{e}{RV members from \citet{terndrup02}, and X-ray detected sources from \citet{damiani03}.}
\end{deluxetable*}

Table~\ref{tab:chi} lists the results from the $\chi^2$ solutions
for various combinations of the data listed in the first column.
The ${\rm [M/H]}_{\chi^2}$ and $E(B - V)_{\chi^2}$ denote our derived
values of metallicity and reddening, respectively, and are
shown in the third and fourth columns. Errors in these quantities are the sizes
of the 1 $\sigma$ contours in Figure~\ref{fig:me}. The $(m - M)_{0,\chi^2}$
in the second column is the average distance from the three CMDs. The errors
in metallicity and reddening were propagated into the error in distance.
We compare these values with previous estimates in the literature
after a discussion of systematic errors in the next section.

\section{SYSTEMATIC ERRORS AND THE ACCURACY OF MS FITTING}

In the previous two sections we showed that MS fitting can be performed
with high internal precision, resulting in errors in distance
moduli of 0.02 mag (i.e., 1\% in distance) at the spectroscopic
metallicity (Table~\ref{tab:dist}). When spectroscopy is not available,
photometry in $BVI_CK_s$ alone can be used to derive distances
with precision of 1\% -- 3\%, metallicities with 0.02 -- 0.04 dex error,
and reddening estimates with better than 0.01 mag error
(Table~\ref{tab:chi}). However, various systematic errors in
the MS fitting should be considered in the final error budget such as
the photometric zero-point errors or biases introduced by the cluster
binaries and foreground/background stars. In this section we estimate
these systematic errors and show that they often exceed the internal
precision of the MS fitting by factors of 2 -- 3.

\subsection{Errors from Input Quantities}\label{sec:sys}

\begin{deluxetable*}{llccccccc}
\tablewidth{0pt}
\tablecaption{MS Fitting Error Budget for Praesepe\label{tab:pr}}
\tablehead{
  \colhead{} &
  \colhead{} &
  \colhead{Adopted {\rm [Fe/H]}} &
  \multicolumn{2}{c}{Adopted $E(B -  V)$} &&
  \multicolumn{3}{c}{$\chi^2$ Minimization} \nl
  \cline{4-5} \cline{7-9}
  \colhead{Source of Error} &
  \colhead{$\Delta$Quantity} &
  \colhead{$\Delta (m -  M)_{0,S}$} &
  \colhead{$\Delta (m -  M)_{0,E}$} &
  \colhead{${\rm [M/H]}_E$} &&
  \colhead{$\Delta (m -  M)_{0,{\chi^2}}$} &
  \colhead{${\rm [M/H]}_{\chi^2}$} &
  \colhead{$E(B -  V)_{\chi^2}$}
}
\startdata
${\rm [Fe/H]}$\dotfill   & $\pm0.03$  & $\pm0.029$ & \nodata    & \nodata    && \nodata    & \nodata    & \nodata    \nl
$E(B -  V)$\dotfill      & $\pm0.002$ & $\pm0.005$ & $\pm0.013$ & $\pm0.007$ && \nodata    & \nodata    & \nodata    \nl
Age\dotfill              & $\pm30\%$  & $\pm0.007$ & $\pm0.010$ & $\pm0.000$ && $\mp0.014$ & $\mp0.013$ & $\mp0.004$ \nl
Helium (Y)\dotfill       & $\pm0.009$ & $\mp0.027$ & $\mp0.027$ & $\pm0.000$ && $\mp0.027$ & $\pm0.000$ & $\pm0.000$ \nl
Calibration\dotfill      &            & $\pm0.010$ & $\pm0.020$ & $\pm0.020$ && $\pm0.010$ & $\pm0.010$ & $\pm0.002$ \nl
Fitting method\dotfill   &            & $\pm0.006$ & $\pm0.014$ & $\pm0.013$ && $\pm0.002$ & $\pm0.004$ & $\pm0.001$ \nl
$R_V$\dotfill            & $\pm0.3$   & $\pm0.000$ & $\pm0.006$ & $\pm0.006$ && $\pm0.000$ & $\mp0.001$ & $\pm0.000$ \nl
$R_{VI}$\dotfill         & $\pm0.07$  & $\pm0.001$ & $\pm0.004$ & $\pm0.003$ && $\mp0.002$ & $\mp0.003$ & $\pm0.001$ \nl
$R_{VK}$\dotfill         & $\pm0.12$  & $\pm0.001$ & $\pm0.002$ & $\pm0.001$ && $\pm0.000$ & $\pm0.000$ & $\pm0.000$ \nl
$\Delta V$\dotfill       & $\pm0.006$ & $\pm0.001$ & $\mp0.011$ & $\mp0.011$ && $\mp0.003$ & $\mp0.008$ & $\pm0.002$ \nl
$\Delta K_s$\dotfill     & $\pm0.007$ & $\pm0.006$ & $\pm0.019$ & $\pm0.013$ && $\pm0.010$ & $\pm0.010$ & $\mp0.002$ \nl
$\Delta (B -  V)$\dotfill   & $\pm0.004$ & $\mp0.006$ & $\pm0.038$ & $\pm0.044$ && $\pm0.014$ & $\pm0.027$ & $\mp0.002$ \nl
$\Delta (V -  I)_C$\dotfill & $\pm0.006$ & $\mp0.008$ & $\mp0.027$ & $\mp0.021$ && $\mp0.007$ & $\mp0.006$ & $\pm0.002$ \nl
Total\dotfill            &            & $\pm0.044$ & $\pm0.065$ & $\pm0.058$ && $\pm0.037$ & $\pm0.035$ & $\pm0.006$ \nl
Fitting\dotfill          &            & $\pm0.021$ & $\pm0.030$ & $\pm0.030$ && $\pm0.025$ & $\pm0.020$ & $\pm0.005$
\enddata
\end{deluxetable*}

\begin{deluxetable*}{llccccccc}
\tablewidth{0pt}
\tablecaption{MS Fitting Error Budget for the Pleiades\label{tab:pl}}
\tablehead{
  \colhead{} &
  \colhead{} &
  \colhead{Adopted {\rm [Fe/H]}} &
  \multicolumn{2}{c}{Adopted $E(B -  V)$} &&
  \multicolumn{3}{c}{$\chi^2$ Minimization} \nl
  \cline{4-5} \cline{7-9}
  \colhead{Source of Error} &
  \colhead{$\Delta$Quantity} &
  \colhead{$\Delta (m -  M)_{0,S}$} &
  \colhead{$\Delta (m -  M)_{0,E}$} &
  \colhead{${\rm [M/H]}_E$} &&
  \colhead{$\Delta (m -  M)_{0,{\chi^2}}$} &
  \colhead{${\rm [M/H]}_{\chi^2}$} &
  \colhead{$E(B -  V)_{\chi^2}$}
}
\startdata
${\rm [Fe/H]}$\dotfill   & $\pm0.02$  & $\pm0.023$ & \nodata    & \nodata    && \nodata    & \nodata    & \nodata    \nl
$E(B -  V)$\dotfill      & $\pm0.003$ & $\pm0.007$ & $\pm0.010$ & $\pm0.002$ && \nodata    & \nodata    & \nodata    \nl
Age\dotfill              & $\pm30\%$  & $  +0.006$ & $  +0.020$ & $  +0.011$ && $  +0.004$ & $  +0.002$ & $  +0.001$ \nl
Helium (Y)\dotfill       & $\pm0.009$ & $\mp0.027$ & $\mp0.027$ & $\pm0.000$ && $\mp0.027$ & $\pm0.000$ & $\pm0.000$ \nl
Calibration\dotfill      &            & $\pm0.010$ & $\pm0.020$ & $\pm0.020$ && $\pm0.010$ & $\pm0.010$ & $\pm0.002$ \nl
Fitting method\dotfill   &            & $\pm0.006$ & $\pm0.038$ & $\pm0.033$ && $\pm0.027$ & $\pm0.014$ & $\pm0.005$ \nl
$R_V$\dotfill            & $\pm0.3$   & $\mp0.003$ & $\pm0.018$ & $\pm0.018$ && $\pm0.013$ & $\pm0.016$ & $\mp0.001$ \nl
$R_{VI}$\dotfill         & $\pm0.07$  & $\pm0.003$ & $\pm0.012$ & $\pm0.008$ && $\pm0.010$ & $\pm0.008$ & $\mp0.001$ \nl
$R_{VK}$\dotfill         & $\pm0.12$  & $\pm0.002$ & $\pm0.008$ & $\pm0.005$ && $\pm0.007$ & $\pm0.005$ & $\pm0.000$ \nl
$\Delta V$\dotfill       & $\pm0.012$ & $\pm0.008$ & $\mp0.009$ & $\mp0.014$ && $\mp0.013$ & $\mp0.018$ & $\pm0.000$ \nl
$\Delta K_s$\dotfill     & $\pm0.007$ & $\pm0.003$ & $\pm0.012$ & $\pm0.008$ && $\pm0.014$ & $\pm0.010$ & $\pm0.000$ \nl
$\Delta (B -  V)$\dotfill   & $\pm0.008$ & $\mp0.023$ & $\pm0.058$ & $\pm0.069$ && $\pm0.048$ & $\pm0.066$ & $\mp0.002$ \nl
$\Delta (V -  I)_C$\dotfill & $\pm0.009$ & $\mp0.013$ & $\mp0.037$ & $\mp0.021$ && $\mp0.034$ & $\mp0.030$ & $\pm0.006$ \nl
Total\dotfill            &            & $\pm0.048$ & $\pm0.092$ & $\pm0.087$ && $\pm0.076$ & $\pm0.080$ & $\pm0.009$ \nl
Fitting\dotfill          &            & $\pm0.013$ & $\pm0.039$ & $\pm0.032$ && $\pm0.053$ & $\pm0.040$ & $\pm0.007$
\enddata
\end{deluxetable*}

\begin{deluxetable*}{llccccccc}
\tablewidth{0pt}
\tablecaption{MS Fitting Error Budget for M67\label{tab:67}}
\tablehead{
  \colhead{} &
  \colhead{} &
  \colhead{Adopted {\rm [Fe/H]}} &
  \multicolumn{2}{c}{Adopted $E(B -  V)$} &&
  \multicolumn{3}{c}{$\chi^2$ Minimization} \nl
  \cline{4-5} \cline{7-9}
  \colhead{Source of Error} &
  \colhead{$\Delta$Quantity} &
  \colhead{$\Delta (m -  M)_{0,S}$} &
  \colhead{$\Delta (m -  M)_{0,E}$} &
  \colhead{${\rm [M/H]}_E$} &&
  \colhead{$\Delta (m -  M)_{0,{\chi^2}}$} &
  \colhead{${\rm [M/H]}_{\chi^2}$} &
  \colhead{$E(B -  V)_{\chi^2}$}
}
\startdata
${\rm [Fe/H]}$\dotfill   & $\pm0.01$  & $\pm0.008$ & \nodata    & \nodata    && \nodata    & \nodata    & \nodata    \nl
$E(B -  V)$\dotfill      & $\pm0.004$ & $\pm0.008$ & $\pm0.016$ & $\pm0.009$ && \nodata    & \nodata    & \nodata    \nl
Age\dotfill              & $\pm30\%$  & $\pm0.006$ & $\pm0.009$ & $\pm0.005$ && $\pm0.001$ & $\mp0.002$ & $\mp0.002$ \nl
Helium (Y)\dotfill       & $\pm0.009$ & $\mp0.027$ & $\mp0.027$ & $\pm0.000$ && $\mp0.027$ & $\pm0.000$ & $\pm0.000$ \nl
Calibration\dotfill      &            & $\pm0.010$ & $\pm0.020$ & $\pm0.020$ && $\pm0.010$ & $\pm0.010$ & $\pm0.002$ \nl
Fitting method\dotfill   &            & $\pm0.004$ & $\pm0.011$ & $\pm0.016$ && $\pm0.001$ & $\pm0.001$ & $\pm0.000$ \nl
$R_V$\dotfill            & $\pm0.3$   & $\mp0.004$ & $\pm0.019$ & $\pm0.026$ && $\pm0.000$ & $\pm0.008$ & $\mp0.002$ \nl
$R_{VI}$\dotfill         & $\pm0.07$  & $\pm0.006$ & $\pm0.014$ & $\pm0.010$ && $\pm0.004$ & $\pm0.002$ & $\mp0.001$ \nl
$R_{VK}$\dotfill         & $\pm0.12$  & $\pm0.001$ & $\pm0.007$ & $\pm0.007$ && $\pm0.002$ & $\pm0.001$ & $\pm0.000$ \nl
$\Delta V$\dotfill       & $\pm0.009$ & $\pm0.006$ & $\mp0.006$ & $\mp0.014$ && $\mp0.007$ & $\mp0.016$ & $\pm0.000$ \nl
$\Delta K_s$\dotfill     & $\pm0.007$ & $\pm0.002$ & $\pm0.011$ & $\pm0.011$ && $\pm0.012$ & $\pm0.012$ & $\pm0.000$ \nl
$\Delta (B -  V)$\dotfill   & $\pm0.005$ & $\mp0.006$ & $\pm0.038$ & $\pm0.049$ && $\pm0.026$ & $\pm0.043$ & $\mp0.003$ \nl
$\Delta (V -  I)_C$\dotfill & $\pm0.011$ & $\mp0.022$ & $\mp0.045$ & $\mp0.023$ && $\mp0.003$ & $\mp0.009$ & $\pm0.014$ \nl
Total\dotfill            &            & $\pm0.040$ & $\pm0.076$ & $\pm0.070$ && $\pm0.041$ & $\pm0.050$ & $\pm0.014$ \nl
Fitting\dotfill          &            & $\pm0.015$ & $\pm0.054$ & $\pm0.062$ && $\pm0.034$ & $\pm0.032$ & $\pm0.006$
\enddata
\tablecomments{Estimated from S04 photometry.}
\end{deluxetable*}

\begin{deluxetable*}{llccccccc}
\tablewidth{0pt}
\tablecaption{MS Fitting Error Budget for NGC~2516\label{tab:25}}
\tablehead{
  \colhead{} &
  \colhead{} &
  \colhead{Adopted {\rm [Fe/H]}} &
  \multicolumn{2}{c}{Adopted $E(B -  V)$} &&
  \multicolumn{3}{c}{$\chi^2$ Minimization} \nl
  \cline{4-5} \cline{7-9}
  \colhead{Source of Error} &
  \colhead{$\Delta$Quantity} &
  \colhead{$\Delta (m -  M)_{0,S}$} &
  \colhead{$\Delta (m -  M)_{0,E}$} &
  \colhead{${\rm [M/H]}_E$} &&
  \colhead{$\Delta (m -  M)_{0,{\chi^2}}$} &
  \colhead{${\rm [M/H]}_{\chi^2}$} &
  \colhead{$E(B -  V)_{\chi^2}$}
}
\startdata
${\rm [Fe/H]}$\dotfill   & $\pm0.07$  & $\pm0.072$ & \nodata    & \nodata    && \nodata    & \nodata    & \nodata    \nl
$E(B -  V)$\dotfill      & $\pm0.002$ & $\pm0.006$ & $\pm0.005$ & $\pm0.000$ && \nodata    & \nodata    & \nodata    \nl
Age\dotfill              & $\pm30\%$  & $\pm0.005$ & $\pm0.014$ & $\pm0.008$ && $\pm0.014$ & $\pm0.006$ & $\pm0.000$ \nl
Helium (Y)\dotfill       & $\pm0.009$ & $\mp0.027$ & $\mp0.027$ & $\pm0.000$ && $\mp0.027$ & $\pm0.000$ & $\pm0.000$ \nl
Calibration\dotfill      &            & $\pm0.010$ & $\pm0.020$ & $\pm0.020$ && $\pm0.010$ & $\pm0.010$ & $\pm0.002$ \nl
Fitting method\dotfill   &            & $\pm0.009$ & $\pm0.016$ & $\pm0.019$ && $\pm0.002$ & $\pm0.003$ & $\pm0.000$ \nl
$R_V$\dotfill            & $\pm0.3$   & $\mp0.009$ & $\pm0.051$ & $\pm0.058$ && $\pm0.030$ & $\pm0.063$ & $\mp0.010$ \nl
$R_{VI}$\dotfill         & $\pm0.07$  & $\pm0.008$ & $\pm0.017$ & $\pm0.009$ && $\pm0.011$ & $\pm0.005$ & $\pm0.000$ \nl
$R_{VK}$\dotfill         & $\pm0.12$  & $\pm0.009$ & $\pm0.031$ & $\pm0.023$ && $\pm0.016$ & $\pm0.021$ & $\mp0.005$ \nl
$\Delta V$\dotfill       & $\pm0.008$ & $\pm0.002$ & $\mp0.011$ & $\mp0.015$ && $\mp0.004$ & $\mp0.013$ & $\pm0.002$ \nl
$\Delta K_s$\dotfill     & $\pm0.007$ & $\pm0.005$ & $\pm0.017$ & $\pm0.013$ && $\pm0.011$ & $\pm0.012$ & $\mp0.002$ \nl
$\Delta (B -  V)$\dotfill   & $\pm0.002$ & $\mp0.004$ & $\pm0.012$ & $\pm0.016$ && $\pm0.011$ & $\pm0.016$ & $\pm0.000$ \nl
$\Delta (V -  I)_C$\dotfill & $\pm0.006$ & $\mp0.007$ & $\mp0.014$ & $\mp0.007$ && $\mp0.009$ & $\mp0.007$ & $\pm0.002$ \nl
Total\dotfill            &            & $\pm0.080$ & $\pm0.079$ & $\pm0.074$ && $\pm0.051$ & $\pm0.072$ & $\pm0.012$ \nl
Fitting\dotfill          &            & $\pm0.023$ & $\pm0.058$ & $\pm0.053$ && $\pm0.038$ & $\pm0.020$ & $\pm0.007$
\enddata
\tablecomments{Estimated from JTH01 photometry.}
\end{deluxetable*}

In Tables~\ref{tab:pr}--\ref{tab:25} we list the sources of
several systematic errors and their contributions to errors in the
distance, photometric metallicity, and reddening estimates for
each cluster. The first column displays the source of errors, and
the second column shows the size of the errors adopted for each
cluster. The third column lists error contributions to $(m - M)_{0,S}$ when
both metallicity and reddening are held fixed. The fourth and fifth columns
contain errors in $(m - M)_{0,E}$ and ${\rm [M/H]}_E$ when the
photometric metallicity is determined at a fixed reddening.
The sixth through eighth columns list errors in $(m - M)_{0,{\chi^2}}$, ${\rm
[M/H]}_{\chi^2}$, and $E(B - V)_{\chi^2}$ from the $\chi^2$
minimization. At the bottom of the table, we list the ``total error,''
computed as the quadrature sum of all of the error
contributions. These are the errors that are shown as error bars
along with the $\chi^2$ contours in Figure~\ref{fig:me}. The fitting
errors (i.e., the internal precision of MS fitting) are not
included in these sums but instead are separately tabulated below.

In the first column of these tables, the line marked ${\rm [Fe/H]}$ denotes
the error in the adopted cluster metallicity (Table~\ref{tab:feh.lit}).
This affects the error in $(m - M)_{0,S}$ only.  The errors in
the adopted reddening (Table~\ref{tab:ebv.lit}) are shown in
the line marked $E(B - V)$. These errors are propagated into
the determination of the photometric metallicity ${\rm [M/H]_E}$ and
the distance $(m - M)_{0,E}$, but not the values determined from
the $\chi^2$ minimization. The remaining errors will contribute to
the systematic errors of all of the parameters found via MS fitting.

We chose an error of 30\% in the adopted age of each cluster from
the consideration of previous age estimates for the Pleiades.
The cluster age is about 100 Myr from the MS turnoff using isochrones
that incorporate convective overshooting \citep{meynet93},
while ages from the lithium depletion boundary are about 125 Myr
\citep{stauffer98,burke04}. However, distances, photometric
metallicities, and reddening values are insensitive to age
because the color range we chose for the fitting avoids the upper MS
and the pre-MS (for clusters at least as old as the Pleiades).

The helium abundance ($Y$) sensitively affects the isochrone luminosity
(${\Delta M_V} / {\Delta Y} \approx 3$ at fixed $T_{\rm eff}$).
A shorter distance is obtained when a higher helium abundance is
assumed. We adopted an error in $Y$ of 0.009 for each cluster. This is
a 1 $\sigma$ scatter in $\Delta Y / \Delta Z$ from the primordial
helium abundance, the solar value, and that of the Hyades as inferred
from the MS luminosity at the {\it Hipparcos} distance (see Paper~I).
The photometric metallicities and reddening values are not affected by
the helium abundance since the error in $Y$ equally changes the distances
in the three CMDs.

The calibration errors in the tables reflect errors in our adopted
parameters of the Hyades: ${\rm [Fe/H]} = +0.13 \pm 0.01$ \citep{paulson03}
and the center-of-mass distance modulus of $(m - M)_0 = 3.33 \pm 0.01$
\citep{perryman98}. We adopted a 1 $\sigma$ error in reddening from
\citet{taylor80}, $E(B - V) = +0.003 \pm 0.002$, but assumed a zero
reddening toward the Hyades (Paper~II).
The errors in ${\rm [M/H]}_E$ and $(m - M)_{0,E}$ further include
the errors arising from the scatter about the linear relation between
the two quantities (see Fig.~\ref{fig:hy.md}).
The error from the fitting method comes from experiments in which we
used the weighted mean or weighted median distance modulus instead of
the unweighted median value.

The next three rows in the tables show the effects of errors in the
reddening laws. \citet{wegner03} determined $R_V$ towards about 600 OB
stars, from which we estimated an uncertainty of $\pm 0.3$ from the MAD
of the distribution. We estimated the dependence of $R_{VI}$ and $R_{VK}$
on the choice of $R_V$ from the extinction law of \citet{cardelli89},
and found $\Delta R_{VI} / \Delta R_V \simeq 0.22$ and
$\Delta R_{VK} / \Delta R_V \simeq 0.84$ for B-type stars.
We adopted $\pm 0.07$ for the error in $R_{VI}$
from the difference between our adopted value and $R_{VI} = 1.24$
from \citet{dean78}. Similarly, we adopted $\pm 0.12$ as the
error in $R_{VK}$ by considering our value with those advocated
by \citet{schultz75} and by \citet{sneden78}. These are
significant contributors to the error budget for the highly-reddened
cluster (NGC~2516), but are much less significant for the other clusters.

The next four rows list the errors resulting from zero point
errors in the photometry. In this calculation, we assumed that
the fundamental observed quantities are $V$, $K_s$, $B - V$, and
$V - I_C$ \citep[e.g.,][]{stetson03}. The adopted values of
the errors are shown in the second column. The error in $K_s$ was
taken as the calibration uncertainty (0.007 mag) that was specified
in the explanatory supplement to the 2MASS All Sky Data
Release.\footnote{See http://www.ipac.caltech.edu/2mass/releases/allsky/doc/explsup.html.}
For the Pleiades and Praesepe, the errors come from intercomparisons
between limited subsets of stars in common among various studies
and from a consideration of the entries in Tables~\ref{tab:phot.diff}
and \ref{tab:tycho.diff}. For M67 and NGC~2516, we assumed
the magnitude and color errors to be half the difference between
the two studies (Table~\ref{tab:phot.diff}).

Besides the systematic errors in the photometry, distance estimation
can be affected by the variability of pre-MS and MS stars
\citep[e.g.,][]{scholz04}, which may be caused by rotational modulation
of cool and hot regions on stellar surfaces. This issue can be
addressed with multiple observations for each star, but those kinds
of data are not usually available.

Errors in the metallicity and the helium abundance are important
contributors to the error in the distance $(m - M)_{0,S}$. As shown
in Table~\ref{tab:feh.lit}, individual spectroscopic studies
generally have $\sim 0.03$ dex errors in the mean metal abundance,
and this is propagated into the error of $\sim 0.03$ mag in distance
modulus since $\Delta (m - M)_0 / \Delta {\rm [Fe/H]} \approx 1$.
In M67, the error in the metallicity has no greater effect than
the other systematic errors due to the many independent metallicity
estimates. This contrasts well with the NGC~2516 case where only
two stars were measured from high-resolution spectroscopy.
The zero point errors in the photometry dominate the systematic
errors in the photometric metallicity and the distance derived
purely from the isochrones, independently of the spectroscopy.
However, our analysis demonstrates that the photometric metallicity
and the distance can be derived with comparable accuracy to those
from the spectroscopic studies. In all cases, the fitting errors
are smaller than the total systematic errors by factors of 2 -- 3
as shown in the tables.

The cluster richness can also affect the distance determination.
If there are $N$ genuine single cluster members with random photometric
errors in colors of $\epsilon$, the error in distance would be
approximately $s \epsilon (3 N)^{-0.5} \approx 2 \epsilon N^{-0.5}$,
where $s$ is a typical slope of MS. This error was already included
in the fitting error, but its contribution is small for clusters
with good photometry and $N \approx 100$. However, information on
single cluster members is only available for a handful of open clusters,
and the effects of binaries and foreground/background stars should be
taken into account in distance estimation, as is discussed in
the next two sections.

\subsection{Cluster Binaries}\label{sec:bin}

The filtering algorithm (\S~\ref{sec:filtering}) removes cluster
binaries if they are sufficiently far from the MS. However, low mass
ratio binaries would remain and might systematically reduce
the estimated distance since they are brighter and redder than single
cluster members.

To determine the size of the bias in distance estimation and to evaluate
the effectiveness of the filtering algorithm, we performed artificial
cluster CMD tests using a solar-metallicity isochrone for an age of
550 Myr. We constructed each set of CMDs by generating 200 single stars
or unresolved binaries in the color range used for the MS fitting,
$0.4 \leq (B - V)_0 \leq 1.3$. Primaries and secondaries in the binaries,
as well as single MS stars, were randomly drawn from the mass function
(MF) for M35 \citep{barrado01}. The lower mass limit for the secondaries
was set to be the hydrogen-burning limit at $0.08 M_\odot$.
Each simulation is characterized by the binary fraction (BF), which is
defined as the number of binaries divided by the total number of systems
in the above fitting range. For example, at BF = 50\%, one-half of
the data points correspond to single stars and the other half represent
unresolved binaries. We neglected multiple systems other than binaries
due to the observed low frequency among solar-type stars
\citep{duquennoy91}. We assigned equal photometric errors of 0.01 mag
in $V$, $B - V$, $V - I_C$, and $V - K_s$ and displaced them from
the isochrone assuming a normal distribution. The artificial CMDs were
then filtered, and the distance was derived in the usual way.

\begin{figure}
\epsscale{1.2}
\plotone{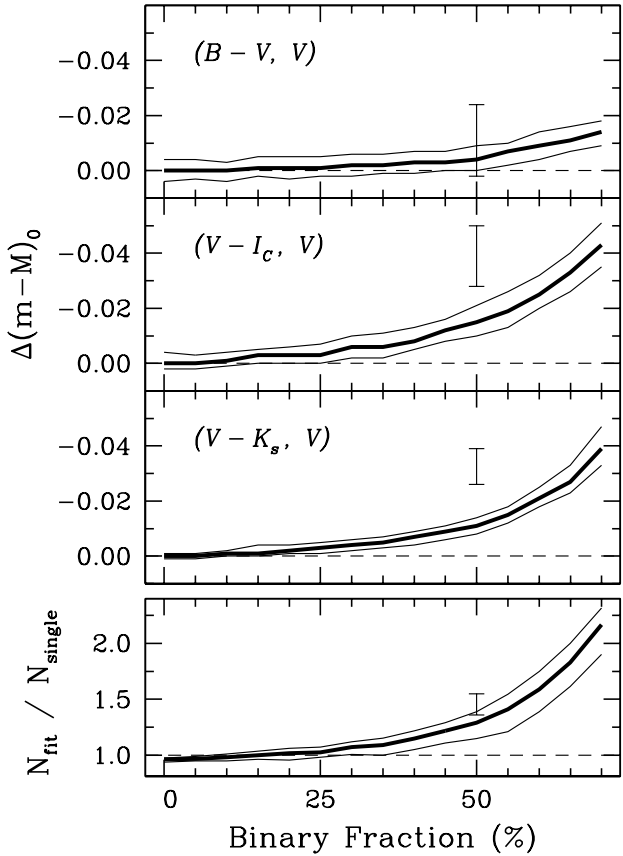}
\caption{Effects of remaining binaries after photometric filtering
from artificial cluster CMD tests (see text for details).
{\it Top three panels:} Bias in distance modulus as a function of binary
fraction in each CMD with assumed photometric errors of 0.01 mag.
{\it Bottom:} Number ratio of all data points used in the fit to
the input single stars. The thick solid line shows the median of these
values from 200 artificial cluster CMDs at each binary fraction with
intervals of 5\%, and thin lines on either side are the first and third
quartiles. The dashed line indicates $\Delta(m - M)_0 = 0$.
The differences are in the sense of shorter distances and more remaining
binaries at a higher binary fraction. The vertical error bar displays
the quartile ranges for a binary fraction of 50\%, but with errors for
each star of 0.03 mag.\label{fig:sim.bin}}
\end{figure}

Figure~\ref{fig:sim.bin} summarizes the results of these simulations.
The top three panels show the bias in the distance modulus as a function
of BF. The thick solid line shows the median of the bias from 200
artificial cluster CMDs computed at each BF with intervals of 5\%. Thin
lines on either side are the first and third quartiles of these
distributions. The dashed line indicates $\Delta(m - M)_0 = 0$.
The vertical error bar displays the quartile ranges for a BF of 50\%,
but with errors for each star of 0.03 mag.

The bottom panel of Figure~\ref{fig:sim.bin} displays the ratio of
the number of data points remaining after filtering compared to the
input number of single stars in the simulation. The lines have the same
meaning as in the other panels. For example, at BF = 50\%,
the simulation yields $N_{\rm fit} / N_{\rm single} \simeq 1.3$ after
filtering, indicating that $\sim70$\% of binaries were eliminated
by the filtering algorithm.

As seen in Figure~\ref{fig:sim.bin}, the distance from the $(B - V, V)$
CMD is less affected by the remaining binaries than those from
$(V - I_C, V)$ and $(V - K_s, V)$. Since stars were rejected if they
were filtered at least in one of the CMDs, the total number
of remaining binaries is the same in the three CMDs with a binary
detection limit set by a certain binary mass ratio\footnote{Binary
mass ratio is defined as the mass of the secondary (less massive) star
divided by the mass of the primary star.} ($\sim0.4$ in our
simulation). However, binaries stand out more prominently in
$(V - I_C, V)$ or $(V - K_s, V)$ than in $(B - V, V)$ because
the addition of the cooler secondaries results in a smaller change in
the combined color in $B - V$ than in the other two color indices.
As also seen in Figure~\ref{fig:sim.bin}, larger photometric errors result
in a bigger bias in distance estimation since binaries are more difficult
to identify when the photometric errors are larger.

\begin{figure}
\epsscale{1.2}
\plotone{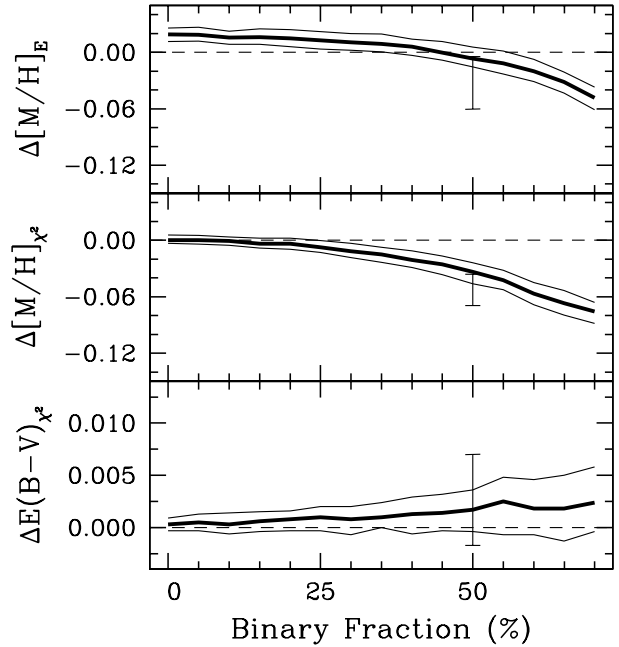}
\caption{Same as Fig.~\ref{fig:sim.bin}, but for the bias in
${\rm [M/H]_E}$, ${\rm [M/H]}_{\chi^2}$, and $E(B - V)_{\chi^2}$
({\it top to bottom}). The differences are in the sense of
a lower metallicity and a higher reddening estimate at a higher binary
fraction.\label{fig:sim.par}}
\end{figure}

The effects of binaries on the determination of the photometric
metallicity and reddening are displayed in Figure~\ref{fig:sim.par}.
The top to bottom panels display the biases in ${\rm [M/H]_E}$,
${\rm [M/H]_{\chi^2}}$, and $E(B - V)_{\chi^2}$, respectively.
Error bars at BF = 50\% show the results with the photometric errors
of 0.03 mag. As the BF increases, the MS fitting gives
lower metallicity estimates. This is a direct result of
the systematic errors in distance as seen in Figure~\ref{fig:sim.bin}.
Recall that if an isochrone at a particular metallicity yields a
distance from $(B - V, V)$ that is longer than from $(V - I_C, V)$
or $(V - K_s, V)$, the photometric metallicity is smaller than
that of the isochrone. Increasing the BF leaves the distances from
$(B - V, V)$ relatively unaltered but gives shorter distances in
the other two colors, thus yielding the lower photometric metallicity.
Nevertheless, the amount of a bias in the photometric metallicity
is $\Delta {\rm [M/H]_{\chi^2}} \approx -0.03$ at BF = 50\%.
In comparison, we conducted some experiments in which the filtering
was not performed and found that this reduced the metallicity
estimates by $\Delta {\rm [M/H]_{\chi^2}} \approx -0.30$.
The results from the bottom panel show that the reddening determination
is almost unaffected by the presence of unfiltered binaries.

In Figure~\ref{fig:sim.par} a small offset of $\Delta {\rm [M/H]}_E
\approx 0.02$ at BF = 0\% shows the systematic error that was already
included in the ``calibration'' error in ${\rm [M/H]}_E$
(Tables~\ref{tab:pr} -- \ref{tab:25}). This happens because
we assumed a linear relation between metallicity and distance over
a wide range of metallicity (e.g., Fig.~\ref{fig:hy.md}); higher order
terms in the relation may be required to eliminate this systematic offset.

\begin{deluxetable*}{lccccc}
\tablewidth{0pt}
\tablecaption{Bias Induced by Unfiltered Binaries from Artificial Cluster Test at BF=50\%\label{tab:bin}}
\tablehead{
  \colhead{MF\tablenotemark{a}} &
  \colhead{$\sigma_{phot}$} &
  \colhead{$\Delta (m -  M)_0$} &
  \colhead{$\Delta {\rm [M/H]}_E$} &
  \colhead{$\Delta {\rm [M/H]}_{\chi^2}$} &
  \colhead{$\Delta E(B -  V)_{\chi^2}$}
}
\startdata
M35     &$\pm0.01$&$-0.010$&$-0.007$&$-0.033$&$+0.002$\nl
M35     &$\pm0.03$&$-0.028$&$-0.034$&$-0.050$&$+0.003$\nl
flat    &$\pm0.01$&$-0.006$&$+0.004$&$-0.020$&$+0.002$\nl
Salpeter\tablenotemark{b}&$\pm0.01$&$-0.010$&$-0.006$&$-0.023$&$+0.000$
\enddata
\tablenotetext{a}{Stellar mass function for secondaries.
Primaries were generated from M35 mass function \citep{barrado01}.}
\tablenotetext{b}{\citet{salpeter55}.}
\end{deluxetable*}

Table~\ref{tab:bin} summarizes the predicted bias in the distance,
photometric metallicity, and reddening induced by unfiltered binaries
at BF = 50\%, which is typical of open clusters \citep{patience02}.
We compare results by adopting different MFs for secondaries, as listed
in the first column. We further tested several combinations of
the various MFs including the \citet{kroupa93} MF for both primaries
and secondaries. However, we found that the detailed form of the MF has
no greater impact on the distance determination than the internal
precision of photometry. The adopted random errors in the photometry are
shown in the second column, while the remaining columns display
the systematic errors.

In Praesepe, many solar-type stars have been observed for binarity
from IR speckle, direct imaging, and spectroscopic studies, covering
a wide range of orbital periods \citep{mermilliod99,bouvier01,patience02}.
As a result, its MS (Figs.~\ref{fig:filter} and \ref{fig:pr.cmd}) is fairly
clean, suggesting that our binarity information in this cluster is
almost complete at least for solar-type stars.
In $0.4 \leq (B - V)_0 \leq 0.8$, there are 46 known binaries out of
116 systems, yielding a BF of 40\%. At this value, our simulation
results in Figures~\ref{fig:sim.bin} and \ref{fig:sim.par} yield
$\Delta (m - M)_{0,S} = -0.006$,
$\Delta {\rm [M/H]}_E = -0.014$, $\Delta {\rm [M/H]}_{\chi^2} = -0.037$,
and $\Delta E(B - V) = +0.003$. These values are consistent with
those from the comparison of two subsamples, computed before and after
excluding known cluster binaries (Tables~\ref{tab:dist} and \ref{tab:chi}).

The Pleiades has also been extensively surveyed on binarity
\citep[e.g.,][]{mermilliod92,bouvier97}. However, there are many stars
remaining above the MS even after excluding all of the known binaries,
suggesting that many binaries were probably left undetected by
the previous surveys (see Fig.~\ref{fig:pl.cmd}). In NGC~2516, there
exists limited information on binarity only for B- and A-type stars from
the spectroscopic studies \citep{abt72,gonzalez00}.

To estimate the BF for each cluster, we first counted the total number
of stars that were 0.3 -- 1.0 mag brighter in $V$ than the cluster MS
in $(B - V, V)$. For NGC~2516, we subtracted field foreground/background
stars based on the distribution of $(m - M)_0$ (see next section) from
the JTH01 catalog before counting likely cluster binaries. Within
the above $\Delta V$ range, the observed BFs are $19 \pm 5\%$ for
the Pleiades and $18 \pm 2\%$ for NGC~2516, while that of Praesepe is
$19 \pm 4\%$. If we combine these estimates with the total observed BF
of 40\% for Praesepe, the total BFs for the Pleiades and NGC~2516 would
be $38 \pm 11\%$ and $38 \pm 4\%$, respectively. The errors
represent those from counting statistics.

Previous studies have also noted that Praesepe and the Pleiades have
the similar BF among G- and K-type stars over a certain range of orbital
periods \citep{mermilliod99,bouvier01,patience02}. However, the above
BF estimates based on the observed fraction of stars with
$\Delta V \geq 0.3$ are uncertain for several reasons. For example,
the distribution of binary mass ratio or the secondary MF can be
different from one cluster to another. If we assume a flat MF for
secondaries, we would derive BF $\approx 50\%$. On the other hand,
the M35 MF predicts quite a high BF ($> 90\%$) and produces too many
binaries near MS (low mass ratio binaries) compared to the observed
distribution of stars. It should also be noted that the BF in
the simulation only concerns photometric binaries. In other words,
blending of physically unrelated stars could increase the BF, while
binaries with large angular separations could reduce the BF.
As a result, the BF depends on a number of cluster properties,
as well as a specific design of a photometric survey.
In addition, a substantial fraction of binaries tend to have
equal-mass companions \citep[e.g.,][and references therein]
{halbwachs03,pinsono06}. These binaries would be well-separated from
the cluster MS, so they can be easily detected and removed by
the filtering algorithm. The net effect would be an overestimation of
the bias in distances and photometric metallicities in the simulation
for a given BF.

From the above considerations, we adopted 30\% -- 50\% as the range of
$\pm 1\ \sigma$ formal errors in the BF for the Pleiades and NGC~2516.
From the simulation result (0.01 mag error, M35 MFs for both binary
components), this BF yields $\Delta (m - M)_S = -0.007 \pm 0.003$,
$\Delta {\rm [M/H]}_E = -0.020 \pm 0.015$,
$\Delta {\rm [M/H]}_{\chi^2} = -0.036 \pm 0.013$, and
$\Delta E(B - V)_{\chi^2} = +0.003 \pm 0.001$.
These values should be subtracted from the MS-fitting results in
Tables~\ref{tab:dist} and \ref{tab:chi} to correct the effects of
binaries. Additional simulations showed that
$\Delta {\rm [M/H]}_{\chi^2} = +0.008 \pm 0.004$ and
$\Delta E(B - V)_{\chi^2} = +0.003 \pm 0.001$ in the case of the fixed
cluster distance. Bias in distance at the photometric metallicity would
be the quadrature sum of $\Delta (m - M)_S$ and the error contributions
from $\Delta {\rm [M/H]}_{\chi^2}$ (or $\Delta {\rm [M/H]}_E$) and
$\Delta E(B - V)_{\chi^2}$.

\subsection{Field Star Contamination}\label{sec:background}

In clusters such as NGC~2516 (Fig.~\ref{fig:25.cmd}), a significant
number of foreground/background stars are present near the MS. Because
the number density of these stars in each CMD increases towards
fainter magnitudes, the distance from MS fitting can be systematically
overestimated.

\begin{figure}
\epsscale{1.2}
\plotone{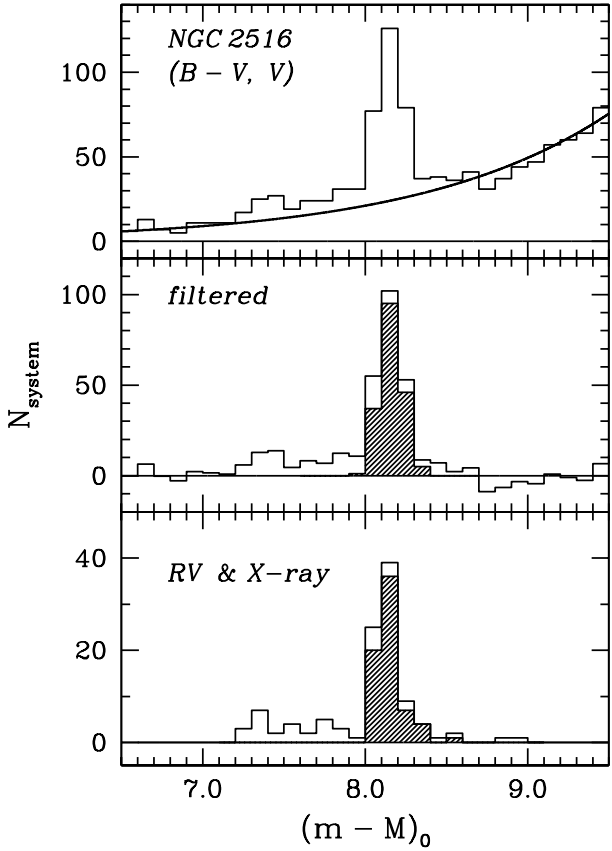}
\caption{Distribution of individual distance moduli for NGC~2516 from
the JTH01 photometry in $(B - V, V)$. Plotted are the number of stars
per 0.1 mag interval in distance modulus. The same isochrone and color
fitting ranges in Fig.~\ref{fig:25.cmd} were used.
{\it Top:} Distribution of the full data in the catalog. The solid line
is an exponential fit to the histogram well beyond the MS and the binary
sequence to model the field star distribution. {\it Middle:} The hatched
histogram represents the distribution of the stars that remained after
filtering. For comparison, the distribution of the full data set after
subtracting backgrounds with the exponential curve in the top panel
is shown as an open histogram. {\it Bottom:} The open histogram is
a summed distribution of the radial velocity members \citep{terndrup02}
and the {\it Chandra} X-ray--detected sources \citep{damiani03},
and the distribution of these samples after filtering is shown as
a hatched histogram.\label{fig:hist}}
\end{figure}

In Figure~\ref{fig:hist} we show the distribution of individual
distance moduli for NGC~2516 from the JTH01 photometry in $(B - V, V)$.
Plotted are the number of stars per 0.1 mag interval in distance modulus.
The same isochrone and color fitting ranges in Figure~\ref{fig:25.cmd}
were used. In the top panel, the histogram represents the number of
stars in the catalog before photometric filtering. It shows the density
peak that represents the cluster MS, the binary sequence that extends
toward shorter distance from the MS, and the field star distribution
that was fitted by an exponential function ({\it solid line}). The fit excluded
stars near the MS and the cluster binaries, and was performed over
a wider range in $(m - M)_0$ than shown here. In the middle panel
the hatched histogram represents the distribution of the stars that
remained after filtering from the catalog. This is compared with an open
histogram, which was found by subtracting the exponential curve from
the histogram of the full data in the top panel.
In the bottom panel we show the summed distribution of the RV members
\citep{terndrup02} and the {\it Chandra} X-ray--detected sources
\citep{damiani03} as an open histogram. The distribution
of these stars after filtering is shown as a hatched histogram.

One of the most conspicuous features in the middle panel of
Figure~\ref{fig:hist} is the effective removal of cluster binaries
from filtering. This can be seen from a deficit of data points in
the hatched histogram against the open one at shorter distances from
the MS. In addition, good matches are found between these two
histograms near the MS, indicating that the filtering algorithm has
worked correctly in the foreground/background subtraction.

The RV and X-ray samples contain many cluster binaries
(Fig.~\ref{fig:hist}, {\it bottom panel}), but they are mostly free from
field star contamination. On the other hand, the full catalog data
(Fig.~\ref{fig:hist}, {\it top panel}) are contaminated by both cluster
binaries and foreground/background stars. Therefore, the effects of
field star contamination after filtering can be estimated by comparing
distances, metallicities, and reddening values derived from each
filtered set of the data (Fig.~\ref{fig:hist}, {\it hatched histograms}).
From Tables~\ref{tab:dist} and \ref{tab:chi}, the weighted mean
differences in these quantities from the JTH01 and S02 photometry are
$\Delta (m - M)_S = +0.010 \pm 0.002$,
$\Delta {\rm [M/H]}_E = +0.020 \pm 0.003$,
$\Delta {\rm [M/H]}_{\chi^2} = +0.038 \pm 0.031$, and
$\Delta E(B - V)_{\chi^2} = -0.003 \pm 0.007$.
The errors are half of the differences in the estimated bias from
the two photometric data sets.  Within the estimated errors, these
values are of the same size, but opposite direction as the binary
corrections at BF = 40\%. In other words, these two biases cancel
out each other.

It is possible that the RV or X-ray samples are more heavily
contaminated by cluster binaries than the full catalog data. However,
we found that they have a similar fraction of stars ($\sim19\%$) in
$0.3 \leq \Delta V \leq 1.0$ mag above the MS. This implies that
the RV and X-ray samples would have a similar total BF unless they have
an excess of low mass ratio binaries.

\subsection{Comparison with Previous Estimates}

\begin{deluxetable*}{lcccc}
\tablewidth{0pt}
\tablecaption{Summary of Open Cluster Parameters\label{tab:final}}
\tablehead{
  \colhead{} &
  \colhead{Praesepe} &
  \colhead{Pleiades} &
  \colhead{M67} &
  \colhead{NGC~2516}
}
\startdata
\multicolumn{5}{c}{{\rm [Fe/H]}} \nl
\hline
This paper\tablenotemark{a} & 
 $+0.20\pm0.04$ & $+0.04\pm0.05$\tablenotemark{b} & $-0.02\pm0.05$\tablenotemark{c} & $-0.07\pm0.06$\tablenotemark{d}\nl
Literature      & 
 $+0.11\pm0.03$ & $+0.04\pm0.02$ & $+0.00\pm0.01$ & $+0.01\pm0.07$\nl
Weighted mean   & 
 $+0.14\pm0.02$ & $+0.04\pm0.02$ & $+0.00\pm0.01$ & $-0.04\pm0.05$\nl
\hline
\multicolumn{5}{c}{$E(B -  V)$} \nl
\hline
This paper\tablenotemark{a} &
 $-0.003\pm0.008$ & $0.022\pm0.014$\tablenotemark{b} & $0.042\pm0.019$\tablenotemark{c} & $0.125\pm0.014$\tablenotemark{d}\nl
Literature      & 
 $0.007\pm0.002$ & $0.032\pm0.003$ & $0.041\pm0.004$ & $0.117\pm0.002$\nl
Weighted mean   & 
 $0.006\pm0.002$ & $0.032\pm0.003$ & $0.041\pm0.004$ & $0.117\pm0.002$\nl
\hline
\multicolumn{5}{c}{$(m -  M)_0$} \nl
\hline
This paper\tablenotemark{e} &
 $6.33\pm0.04$ & $5.63\pm0.02$\tablenotemark{f} & $9.61\pm0.03$\tablenotemark{c} & $8.03\pm0.04$\tablenotemark{d}\nl
\citet{robichon99} &
 $6.28\pm0.13$ & $5.36\pm0.06$ & \nodata & $7.70\pm0.16$\nl
\citet{vanleeuwen99} &
 $6.37\pm0.15$ & $5.37\pm0.07$ & \nodata & \nodata
\enddata
\tablenotetext{a}{Solution from the $\chi^2$ minimization.}
\tablenotetext{b}{Solution at the average geometric distance (Table~\ref{tab:hipp}).}
\tablenotetext{c}{Average solution from MMJ93 and S04 photometry with single cluster membership by S04.}
\tablenotetext{d}{Average solution from JTH01 and S02 photometry.}
\tablenotetext{e}{Estimated at the weighted mean {\rm [Fe/H]} and $E(B -  V)$.}
\tablenotetext{f}{Average geometric distance from Table~\ref{tab:hipp}.}
\end{deluxetable*}

In Table~\ref{tab:final} we summarize our metallicity and reddening
estimates from the $\chi^2$ minimization, and compare them with
the literature values. The {\rm [M/H]} and $E(B - V)$ for the Pleiades
are those derived at the average geometric distance in
Table~\ref{tab:hipp}. These estimates for the Pleiades were also
corrected for the bias induced by the binary contamination. This
correction was applied to the result from the data set that originally
contained all known binaries. For NGC~2516, we did not apply the bias
correction for both binaries and field star contamination for the reason
explained in the previous section.
The errors are the quadrature sum of the fitting and total systematic
errors in Tables~\ref{tab:pr} -- \ref{tab:25}. For the Pleiades and
NGC~2516, we additionally include the errors in the bias from binaries.
Our estimates for M67 and NGC~2516 are the weighted averages from
the two photometric data sets for each cluster. We adopted half of the difference
in these estimates as the 1 $\sigma$ error in the case where they do not
agree with each other within the estimated errors.
In the third row of each quantity, weighted averages of our estimates
and those in the literature are shown; however, the averages remain
close to the literature values in most cases.
Our estimated distances at these average metallicities and reddening
values are shown and compared with the {\it Hipparcos} measurements
\citep{robichon99,vanleeuwen99} in the last three rows.
We discuss the {\it Hipparcos} measurements in \S~\ref{sec:hipp}.

\begin{figure}
\epsscale{1.2}
\plotone{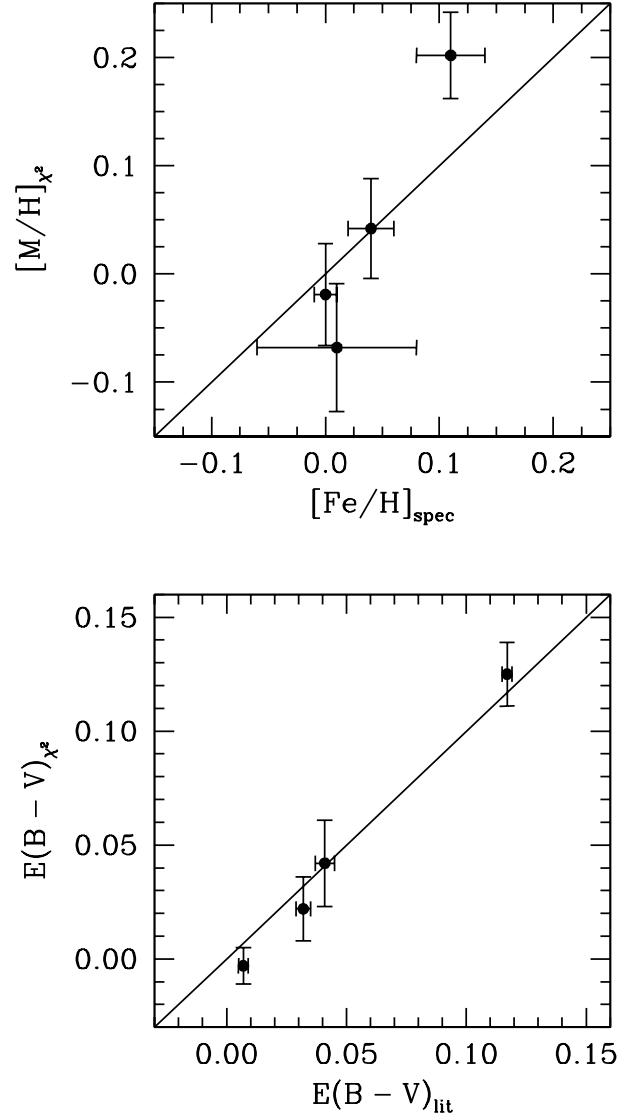}
\caption{Comparison of the photometric ${\rm [M/H]}_{\chi^2}$
and the spectroscopic {\rm [Fe/H]} ({\it top}) and that of
the photometrically determined $E(B - V)_{\chi^2}$ and
the literature value ({\it bottom}) from Table~\ref{tab:final}.
The diagonal lines indicate identity.\label{fig:par.comp}}
\end{figure}

In the top panel of Figure~\ref{fig:par.comp} we compare the
photometric metallicities with the spectroscopic values in
Table~\ref{tab:final}. The comparison yields
$\langle {\rm [M/H]}_{\chi^2} - {\rm [Fe/H]}_{\rm spec} \rangle =
+0.01 \pm 0.03~{\rm (s.e.m.)}$.
Although our photometric metallicity for Praesepe is $\sim 2\ \sigma$
higher than the spectroscopic value, the $\chi^2$ with respect to
the identity line is 4.3 from these four cluster points. In other words,
the errors in ${\rm [M/H]}_{\chi^2}$ were correctly estimated.
We investigate a photometric metallicity scale over a wider range of
{\rm [Fe/H]} in the next paper of this series
(D.\ M.\ Terndrup et al.\ 2007, in preparation).

The $E(B - V)$ estimates from the $\chi^2$ minimization are plotted
against the literature values in the bottom panel of
Figure~\ref{fig:par.comp}. Here the mean difference between the two
measures is $\langle E(B - V)_{\chi^2} - E(B - V)_{\rm lit} \rangle
= -0.006 \pm 0.012~{\rm (s.e.m.)}$. The $\chi^2$ is 2.4, which
again indicates that the errors were estimated correctly. However,
our photometric reddening from $BVI_CK_s$ data is not
completely independent from the literature values from $UBV$ data
since most of the literature estimates in Table~\ref{tab:ebv.lit} are
from the similar methodology that uses the observed colors of
the unreddened MS stars. We did not include $U$-band measurement in
the isochrone calibration because of its low reproducibility and
relatively large systematic errors among different studies
\citep[][and references therein]{bessell05}. These properties make it
difficult to reliably use the $U$-band data in the high-precision
distance estimates as in this paper.

\section{OTHER ISSUES}

\subsection{Are the K Dwarfs in NGC~2516 Blue, Too?}\label{sec:activity}

\begin{figure}
\epsscale{1.2}
\plotone{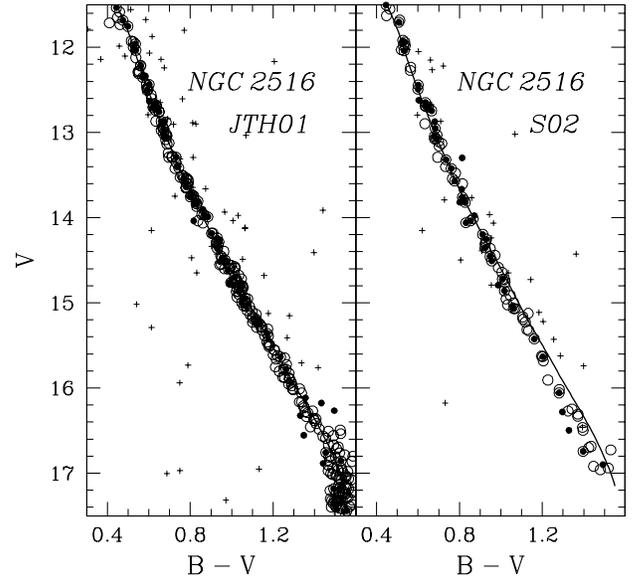}
\caption{Two different interpretations on young K dwarfs in NGC~2516.
Photometry of JTH01 ({\it left}) and that of S02 ({\it right}) are
shown. Open circles are the stars that remained after photometric
filtering from the full catalogs. Filled circles and plus signs show
the stars that are RV members \citep{terndrup02} or X-ray--detected
sources \citep{damiani03} if they were accepted or rejected, respectively,
by the filtering on these samples. The same isochrone and
color fitting ranges in Fig.~\ref{fig:25.cmd} were used. Note that
stars with the S02 photometry are bluer than the isochrone
for $B - V \ga 1.0$.\label{fig:compcmd}}
\end{figure}

In Figure~\ref{fig:compcmd} we display the MS of NGC~2516 in
$(B - V, V)$, which at an age of 140 Myr may exhibit the phenomenon of
blue K dwarfs as seen in the Pleiades CMD (Stauffer et~al. 2003;
see also Fig.~\ref{fig:pl.cmd}).
The left panel displays the photometry of JTH01 in $(B - V, V)$ while
the right panel shows that of S02. In each panel, the open circles
are those stars that remained after photometric filtering. The filled
circles and plus signs show the stars that are RV members \citep{terndrup02}
or X-ray--detected sources \citep{damiani03} if they were accepted or
rejected, respectively, by the filtering.

The solid lines in Figure~\ref{fig:compcmd} are 140 Myr isochrones with
${\rm [Fe/H]} = +0.01$ and $E(B - V) = 0.117$, fitted over
$0.40 \leq (B - V)_0 \leq 0.80$. The isochrone matches the JTH01
photometry down to at least $V \approx 17$, but the SBBL02 photometry
is bluer than the isochrone for $B - V \ga 1.0$ ($M_V \ga 6.5$).
This systematic difference was previously noted when we compared
the photometry from the two studies (Fig.~\ref{fig:25.compv}).

\begin{figure}
\epsscale{1.2}
\plotone{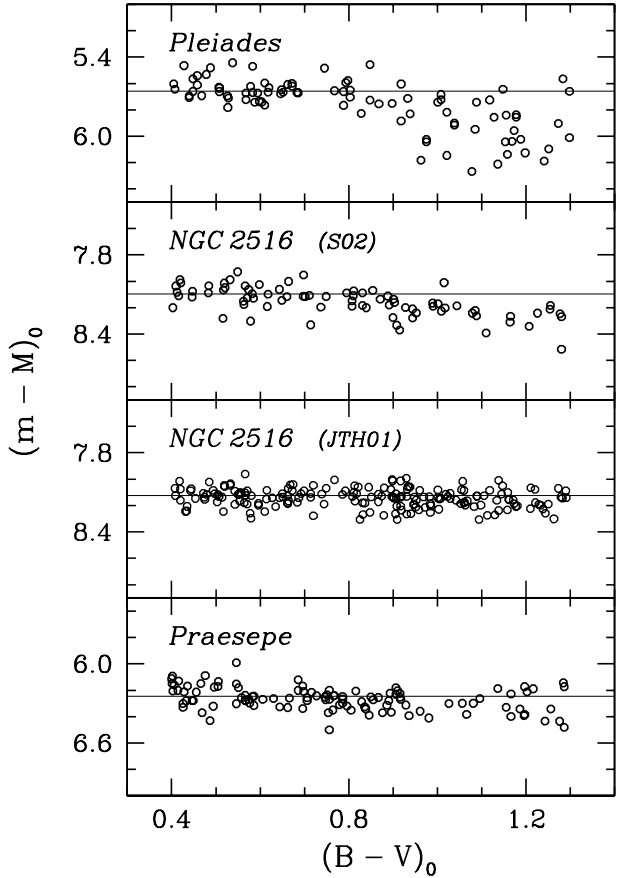}
\caption{Individual distance moduli of stars in the Pleiades
({\it top}), NGC~2516 ({\it middle}), and Praesepe ({\it bottom})
from $(B - V, V)$ with respect to the isochrones with spectroscopic
metallicities and reddening values in the literature
(Tables~\ref{tab:feh.lit}, \ref{tab:ebv.lit}). Only stars that remained
after filtering are shown. Horizontal lines indicate median distance
moduli derived in $0.4 \le (B - V)_0 \le 0.8$. While the JTH01 data
show a flat distribution, the S02 data exhibit longer distances at
redder colors where the Pleiades K dwarfs also depart from
the isochrone.\label{fig:res}}
\end{figure}

The individual distance modulus for each star is shown in
Figure~\ref{fig:res}, plotted against $(B - V)_0$, along with
the Pleiades and Praesepe stars that have similar colors. The solid line
in each panel is the median value from fits over
$0.4 \leq (B - V)_0 \leq 0.8$ (note that our analysis above used
the full color range for Praesepe). While the JTH01 data show a flat
distribution over $0.4 \leq (B - V)_0 \leq 1.3$ as in Praesepe, the red
stars in the S02 data systematically trend toward longer distances
for $(B - V)_0 \ga 0.8$ where the Pleiades K dwarfs also depart from
the isochrone.

\begin{figure}
\epsscale{1.2}
\plotone{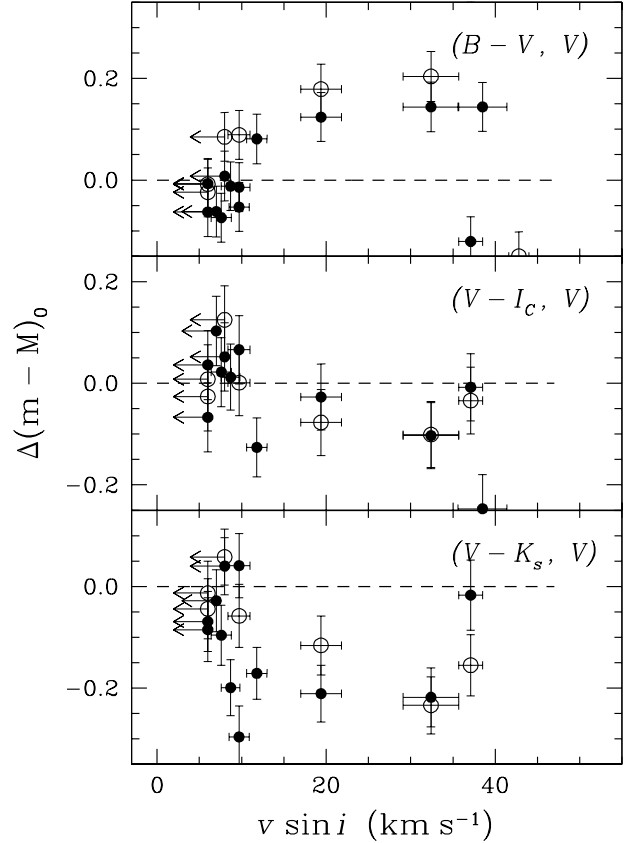}
\caption{Dependence of the photometric anomalies of K dwarfs in NGC~2516
on the projected rotational velocity. The filled circles are from
the JTH01 photometry, and the open circles are from S02 for
$0.8 \leq (B - V)_0 \leq 1.0$. Displayed are the individual distance
moduli with respect to the average distance in each CMD, and
the dashed line shows $\Delta (m - M)_0 = 0$. The same isochrone and
color fitting ranges in Fig.~\ref{fig:25.cmd} were used. Positive
$\Delta (m - M)_0$ indicates that stars are fainter than the isochrone.
\label{fig:vsini}}
\end{figure}

From these plots alone it is not possible to know whether the blue
K dwarf phenomenon exists in NGC~2516 since the interpretation
depends on which data set is used. \citet{stauffer03} argued that the
departure from the ordinary MS in the Pleiades is connected with stellar
rotation rate since the stars with the fastest rotation show the most
anomalous colors (see their Fig.~15). In Figure~\ref{fig:vsini}
we display the difference in $(m - M)_0$ for stars in NGC~2516 from
the average distance modulus in each CMD against the projected rotation
velocity. The filled circles are from the JTH01 photometry, while
the open circles are from the S02. Positive $\Delta (m - M)_0$
indicates that stars are fainter (or bluer) than the isochrone, and
the dashed line shows $\Delta (m - M)_0 = 0$. The $v \sin i$ data are
only available up to $V \approx 15$ or $(B - V)_0 \la 1.0$
\citep{terndrup02}, and only stars with $(B - V)_0 \geq 0.8$ are shown.
Within this restricted color range, a few stars with rapid rotation in
$(B - V, V)$ are about 0.1 -- 0.2 mag fainter than the slow rotators.
By comparison, rapidly rotating stars with similar colors in
the Pleiades are about 0.5 mag below the MS. This is suggestive that
the stars in NGC~2516 may also exhibit the blue K dwarf phenomenon
although the degree of departure is smaller than in the Pleiades.
The opposite correlation is seen in $(V - K_s, V)$ as also noted
by \citet{stauffer03}, and we see no significant correlation in
$(V - I_C, V)$.

\subsection{{\it Hipparcos} Distances}\label{sec:hipp}

\begin{figure}
\epsscale{1.2}
\plotone{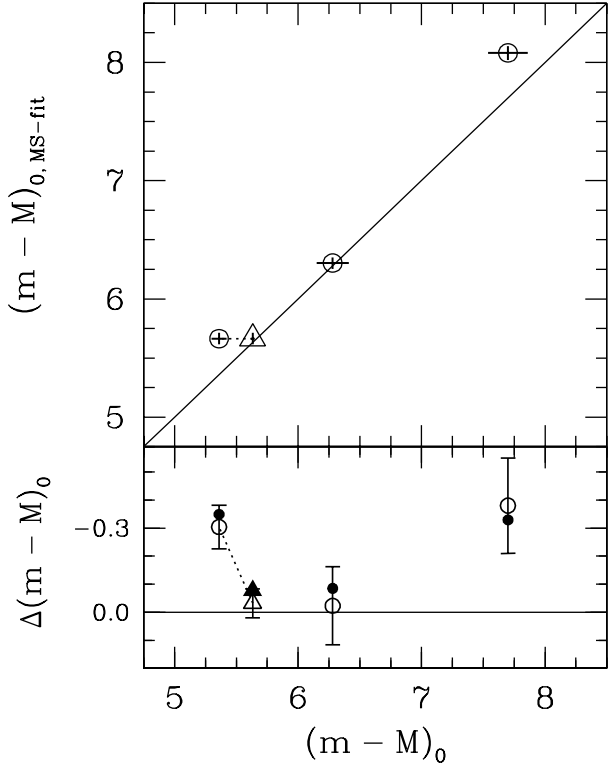}
\caption{Comparison of the {\it Hipparcos} parallax results \citep{robichon99}
and our derived distance moduli. Open circles are those for the Pleiades,
Praesepe, and NGC~2516 ({\it left to right}) when distances are determined
at the spectroscopic metallicities. The open triangles are the comparison
with the geometric distance to the Pleiades (Table~\ref{tab:hipp}), and
the two Pleiades points are connected with a dashed line. The errors are
the quadrature sum of the fitting and total systematic errors in
Tables~\ref{tab:pr} -- \ref{tab:25} and those in the bias induced by
binaries. Filled symbols are the comparison with $(m - M)_{0,\chi^2}$
and are shown without error bars. The diagonal line indicates identity
in the top panel.\label{fig:dist}}
\end{figure}

In Figure~\ref{fig:dist} we compare our derived distance moduli with
the {\it Hipparcos} parallax results from \citet{robichon99} for
the Pleiades, Praesepe, and NGC~2516 (M67 was too distant for {\it Hipparcos}
to measure its parallax). In the top panel the open circles show
the comparison with $(m - M)_{0,S}$ at the spectroscopic metallicity.
The errors are the quadrature sum of the fitting and total systematic
errors in Tables~\ref{tab:pr} -- \ref{tab:25}. For the Pleiades and
NGC~2516, we additionally included the error in the bias correction
for binaries. We also plot as an open triangle the comparison of
$(m - M)_{0,S}$ and the geometric distance to the Pleiades from
Table~\ref{tab:hipp} and connect the two Pleiades points with a dashed
line. The bottom panel shows the difference between the two distance
measurements. The filled circles represent the comparison with
$(m - M)_{0,\chi^2}$ and are shown without error bars.

While our distance to Praesepe shows a good agreement with the {\it Hipparcos}
value, the 1999 {\it Hipparcos} distance to the Pleiades exhibits more than
a $\sim 3\ \sigma$ difference at either the spectroscopic or photometric
metallicity even if we take into account all systematic errors.
Similarly, NGC~2516 shows a $\sim 2\ \sigma$ difference. These
results extend the previous discussion by \citet{pinsono98}
and \citet{terndrup02}, but the current results are based on isochrones
with improved color calibrations over wider color ranges and with
additional $(V - K_s)$ colors.

In fact, \citet{pinsono00} derived a consistent distance to NGC~2516
with the {\it Hipparcos} value, but with a considerably lower photometric
metallicity (${\rm [M/H]} = -0.26$) than presented in this paper.
This happened because $B - V$ and $V - I_C$ colors of their isochrones
from the Yale color calibration \citep{green88} were affected by
large systematic errors in the color-$T_{\rm eff}$ relations.
The photometric metallicity was redetermined to be ${\rm [M/H]} = -0.05$
by \citet{terndrup02}, who employed the same colors, but with
a preliminary calibration of the isochrones that later appeared in our
Paper~II. As in this paper, they found a distance to NGC~2516 that was
longer than the {\it Hipparcos} value.

In the Pleiades, we achieved an excellent fit in all CMDs and derived
a photometric metallicity and distance that are consistent with
the spectroscopic determinations and the geometric distances from
a variety of methods. In our analysis, we excluded $B - V$ colors where
the cluster K dwarfs are anomalously blue. \citet{percival03} presented
$BV(RI)_C$ photometry for a sample of local stars with accurate
{\it Hipparcos} parallaxes and developed a model-independent MS-fitting
method based on the relation between luminosity and metallicity for
these stars. In order to select unevolved stars, however, they picked
targets with $0.7 \la (B - V) \la 1.0$ and found a Pleiades distance
in $(B - V, V)$ that was longer than that from $(V - I_C, V)$,
indicating a photometric metallicity of ${\rm [M/H]} \approx -0.3$. This
led them to suggest that the spectroscopic metallicity of the Pleiades
was in error, even though it is well established
(Table~\ref{tab:feh.lit}). As they subsequently realized
\citep{percival05}, their analysis was seriously affected by
the anomalous $B - V$ colors of the K dwarfs. In the later study, they
showed that distances derived with $V - I_C$ and $V - K_s$ were
consistent with the correct distance for the Pleiades.

The disagreement with the {\it Hipparcos} values for the Pleiades and NGC~2516
has been attributed to a local zero point error of the individual
stellar parallaxes in the {\it Hipparcos} catalog \citep{pinsono98,vanleeuwen05a}.
A more elaborate reduction of the {\it Hipparcos} data \citep{vanleeuwen05b}
would help to resolve the difference from the MS-fitting distances.

\subsection{The Helium Abundance of the Pleiades}\label{sec:helium}

Because the luminosity of the model isochrones at fixed
$T_{\rm eff}$ is sensitive to the helium abundance
($\partial M_{\rm bol} / \partial Y \approx 3$), we can derive
the helium abundance if the cluster distance and metallicity are known.
The isochrones employed here assume that the Sun's metal abundance is
$Z_\odot = 0.0176$ and that the solar helium abundance in models
lacking diffusion is $Y_\odot = 0.266$. This yields
$\Delta Y / \Delta Z = 1.2$ with the primordial helium abundance of
$Y_p = 0.245 \pm 0.002$ (see Paper~I).\footnote{We assumed that
the helium abundance is a function of the heavy-element content $Z$
as given by $Y = Y_p + (\Delta Y / \Delta Z) Z$.} We chose the solar
models with no diffusion because microscopic diffusion would not be
important for young clusters such as the Hyades and the Pleiades.
In Paper~I we derived the Hyades helium abundance,
$Y_{\rm Hyades} = 0.271 \pm 0.006$, using the {\it Hipparcos}
distance to the Hyades \citep{debruijne01}, combined with
the metallicity estimate from \citet{paulson03} and the masses of
the components of the eclipsing binary vB~22 \citep{torres02}. Since
the standard solar model including the effects of helium diffusion and
rotational mixing implies an initial solar helium abundance of
$Y_{\odot,\rm init} = 0.274$ \citep{bahcall01}, the Hyades abundance is
equal within the errors to the Sun's value at the time of formation,
even though the Hyades has $\sim35\%$ higher metal content than the Sun.

At the spectroscopic metallicity for the Pleiades,
${\rm [Fe/H]} = +0.04 \pm 0.03$, we derived
$(m - M)_{0,S} = 5.664 \pm 0.041$, where the error is the quadrature
sum of the fitting and total systematic errors from Table~\ref{tab:pl},
excluding the error contribution from helium. It also includes the error
in the bias correction for binaries. Our distance is longer
than the geometric distance, $(m - M)_0 = 5.632 \pm 0.017$
(Table~\ref{tab:hipp}), which indicates that the helium abundance of
the Pleiades is higher than the assumed value in the model.
Since the isochrones have $Y = 0.268$ at the metallicity of
the Pleiades, the difference between the two distances yields
$Y = 0.279 \pm 0.015$ for the Pleiades.

The Pleiades helium abundance is the same as both the initial solar
and the Hyades' abundances within the errors. To know whether there is
a dispersion in the local helium abundance, it would be necessary
to reduce the size of the systematic errors in the MS fitting
either by improving the photometric calibration or by obtaining
more accurate cluster metal abundances. However, we can certainly
rule out extreme helium abundances $(Y \approx 0.34)$ that had been
advocated to explain the shorter {\it Hipparcos} distance to the Pleiades
\citep{belikov98}. From the primordial helium abundance, the initial
solar value, and the helium abundances for the Hyades and the Pleiades,
we derive $\Delta Y / \Delta Z = 1.42 \pm 0.17$.

When future astrometric missions such as {\it Gaia} \citep{perryman01} provide
highly accurate distances, the error in the helium abundance will be
predominantly determined by the accuracy of the MS fitting. In such
cases, cosmic scatter of $\Delta Y = 0.02$ could be detectable with
the current accuracy of the MS-fitting technique. Furthermore, the error
in the helium enrichment parameter scales as
$\epsilon{(\Delta Y/\Delta Z)} \approx 0.3(N/100)^{-0.5}
(\sigma_{\mu} / 0.05\ {\rm mag})$, where $N$ is the total number of
clusters and $\sigma_{\mu}$ is the error in the distance from
the MS fitting. A more accurate value of $\Delta Y/\Delta Z$ requires
that many input systematic errors in the MS fitting be kept under
control. Stellar abundances in particular are susceptible to systematic
errors from the adopted $T_{\rm eff}$ scales. This issue will be
addressed in the next paper of this series
(D.\ M.\ Terndrup et al.\ 2007, in preparation).

\section{SUMMARY AND DISCUSSION}

We have generated a set of isochrones near solar metallicity
($-0.3 \leq {\rm [Fe/H]} \leq +0.2$ in 0.1 dex increments) and have
calibrated them by correcting the \citet{lejeune97,lejeune98}
color-$T_{\rm eff}$ relations to give a precise match to the Hyades MS
at ${\rm [Fe/H]} = +0.13 \pm 0.01$ and $(m - M)_0 = 3.33 \pm 0.01$
(Paper~II). We tested the hypothesis that the empirical
color-$T_{\rm eff}$ corrections improve the accuracy of distance
estimation using the four well-studied open clusters -- Praesepe,
the Pleiades, M67, and NGC~2516.

Either by adopting cluster metallicities and reddening values in
the literature or by constraining them from the multicolor CMDs,
we showed that the calibrated set of isochrones not only matches the
cluster MS precisely but also gives internally consistent distances
from $(B - V, V)$, $(V - I_C, V)$, and $(V - K_s, V)$ CMDs.
The internal precision of our derived distances is about a factor
of 5 better than the case without the color calibrations.
The current result supports our previous assertion (Paper~II) that the
largest source of isochrone mismatches stems mainly from the conversion
of theoretically predicted quantities into magnitudes and colors.

We also found that our photometric metallicities are in good agreement
with spectroscopic estimates although the largest difference was found
for Praesepe, where the photometric metallicity was $\sim 2\ \sigma$
higher than the spectroscopic value. In addition, we found
an excellent agreement for the Pleiades between our derived
distance and the geometric determinations other than the original
{\it Hipparcos} measurements, confirming the validity of the empirically
calibrated isochrones.

We explored various systematic errors in distance, photometric
metallicity, and reddening estimates. The distance from MS fitting is
sensitive to the metallicity and helium abundance
but can be estimated to an accuracy of $\pm 1\%$ (fitting error) and
$\pm 2\%$ (systematic error) if there exists a good spectroscopic
abundance measurement. When we determine cluster metallicity and/or
reddening with our isochrones alone, the distance becomes sensitive to
the size of the systematic zero-point errors in the photometry. With
reasonable sizes of the zero-point errors, however, distances can be
determined with comparable accuracy to those at the spectroscopic
metallicities.

We showed that the Pleiades K dwarfs are bluer than the calibrated
isochrone in $(B - V, V)$, confirming the previous study by
\citet{stauffer03}. We further showed that the red stars in NGC~2516
are bluer than the isochrone when the S02 data are used; however,
the JTH01 data closely match the isochrone down to the limit of
the isochrone calibration. Using a limited set of $v \sin i$ data, we
found that a few stars with high $v \sin i$ are fainter by 0.1 -- 0.2 mag
than the slow rotators in $0.8 \leq (B - V)_0 \leq 1.0$. This suggests
that the K dwarfs in NGC~2516 may also exhibit the blue K dwarf
phenomenon, although the degree of departure is smaller than in
the Pleiades. More $v \sin i$ data for lower mass stars and careful
photometric standardization are required to set the upper age limit of
the anomalous colors of active K dwarfs. For young open clusters,
careful attention should be given to this kind of anomaly to avoid
erroneous results in distance.

We also derived a helium abundance for the Pleiades of
$Y = 0.279 \pm 0.015$, based on the average geometric distance to
the Pleiades (Table~\ref{tab:hipp}). This value is equal within
the errors to the Sun's initial helium abundance and that of the Hyades.
This result implies that there is no observed cosmic scatter in
the helium abundance at the level of the accuracy in the current
estimation. To decrease the size of the error, more accurate abundances
for the cluster and smaller systematic errors in the photometry are
required.

There are distinct limits to the current calibration. The dependence of
the MS luminosity and colors on metallicity cannot be inferred from one
calibrating cluster alone and must be either assumed to be correct in
the models (as in this paper) or else checked against local field stars
with known distances and metallicity (D.\ M.\ Terndrup et al.\ 2007,
in preparation). The {\it Hipparcos} data in the Hyades are only complete to
$V \approx 11$ ($M_V \approx 8$), so the empirical calibration in
Paper~II extends only down to $\sim 0.7 M_\odot$. In addition,
the Hyades has an age of about 550 Myr, so younger clusters should be
used to calibrate the upper MS for stars more massive than the Hyades
turnoff stars. All of these issues will be addressed with photometry of
low-mass members in Praesepe and $\alpha$ Persei in the later papers of
this series.

\acknowledgements

We are greatly indebted to Jennifer Johnson for her detailed and
thoughtful reading of the first draft of this paper. Many
suggestions by Constantine Deliyannis are gratefully acknowledged. We
also thank Andrew Gould for useful discussions on the proper motion of
cluster members and Hwankyung Sung for detailed description on his
photometry.
The work reported here was partially supported by grants AST-0206008
and AST-0205789 from the National Science Foundation to the Ohio
State Research Foundation and with funds provided by the Ohio State
University Department of Astronomy.
D.~B.~P.\ acknowledges support from the NASA Postdoctoral Program,
which is administered by the Oak Ridge Associated Universities.
This publication makes use of data products from the Two Micron All Sky
Survey, which is a joint project of the University of Massachusetts
and the Infrared Processing and Analysis Center/California Institute of
Technology, funded by the National Aeronautics and Space Administration
and the National Science Foundation.

\appendix

\section{THE HYADES PHOTOMETRIC CALIBRATION REVISITED}

In \S~\ref{sec:sys} we estimated the systematic errors in distance,
photometric metallicity, and reddening that would result from zero-point
errors in the photometry of target clusters. Here we elaborate
on the systematic errors in the Hyades photometry, which determine
the accuracy of the calibration itself.

In our assembly of the Hyades photometry in Paper~II, we employed
the \citet{bessell79} transformation between the Johnson and
Cousins $V - I$ systems, namely,
\begin{equation}
 (V - I)_C = 0.778 (V - I)_J.\label{eq:jtrans}
\end{equation}
This calibration has a small zero-point error, in that this
produces colors that are somewhat too blue. \citet{cousins80}, for
example, derived
\begin{equation}
 (V - I)_C = 0.77 (V - I)_J + 0.01.\label{eq:cousinstrans}
\end{equation}
As part of a detailed study of local field stars, we have assembled
a large sample of stars with both Cousins and Johnson photometry
(D.\ M.\ Terndrup et al.\ 2007, in preparation). These indicate
\begin{equation}
(V - I)_C = 0.779 (V - I)_J + 0.014.\label{eq:ourtrans}
\end{equation}
Although the size of the zero point is small, it produces
a systematic error in distance of $\sim0.1$ mag. This is
because the Johnson data for the Hyades mostly have $M_V \leq 6$
where the average slope of the MS is ${\Delta V} / {\Delta(V - I)_C}
\approx 7$.

\begin{deluxetable*}{lcr}
\tablewidth{0pt}
\tablecaption{Comparison of Johnson and Kron Transformations\label{tab:vidiff}}
\tablehead{
\colhead{Cluster} &
\colhead{$\Delta(V -  I)_C$\tablenotemark{a}} &
\colhead{$N$\tablenotemark{b}}
}
\startdata
Hyades    & $-0.007 \pm 0.007$ & 5 \nl
Praesepe  & $-0.009 \pm 0.003$ & 7 \nl
Pleiades  & $-0.010 \pm 0.004$ & 27\nl
\enddata
\tablenotetext{a}{Transformed Johnson minus Kron photometry
from previous transformations employed in Paper~II (see text).}
\tablenotetext{b}{Number of stars used in the comparison.}
\end{deluxetable*}

As summarized in Table~\ref{tab:vidiff}, there are a few stars in
the Hyades, Praesepe, and the Pleiades that have $V - I$ photometry on
both the Johnson and Kron systems. These provide additional evidence
that a small redward shift in the transformation of the Johnson colors
is needed. The second column shows the mean magnitude difference in the sense of
Johnson minus Kron after the former were placed on the Cousins system
using the (incorrect) transformation in equation~(\ref{eq:jtrans}). The Kron
colors were transformed to the Cousins system using the cubic polynomial
derived by \citet{bessell87}. Errors are the standard error of the mean
difference, and the number of stars used in the comparison is shown
in the last column. In all cases, the Johnson-to-Cousins transformation
from equation~(\ref{eq:jtrans}) employed in Paper~II produces colors that are
slightly too blue compared to those from the Kron system.

The adoption of the new Johnson-to-Cousins transformation
(eq.~[\ref{eq:ourtrans}]) modifies our Hyades calibration, making
the Hyades empirical isochrone from Paper~II slightly redder in the upper
MS since the photometry for these stars is mainly on the Johnson
system. The net effect is to make our color-$T_{\rm eff}$
correction table (Table~2 in Paper~II) redder by
$\Delta (V - I)_C = 0.012$ for $M_V \leq 6$. The isochrones used
in this paper included this correction, no adjustment for
$M_V \geq 8$, and a linear transition between these two
magnitude limits. We also employed equation~(\ref{eq:ourtrans}) in
the assembly of the Praesepe and the Pleiades photometry in this
paper.

In Paper~II, we did not include $V - I_C$ from \citet{taylor85}
because we were unable to achieve a good transformation between
their photometry and those on the Johnson system from other
sources \citep{johnson66,mendoza67,johnson68}. The exclusion of
their data prompted \citet{taylor05} to discuss their photometry
at length and to examine whether there were systematic errors in
the source photometry used in our previous work. They first
verified that their $V - I_C$ values are about 0.02 mag redder than
ours. Then, they identified two sources of the difference:
the adjustment in the Johnson-Cousins transformation noted
above (but derived independently here), and a systematic error in
Mendoza's $R - I_J$ photometry.

\begin{figure}
\epsscale{0.95}
\plotone{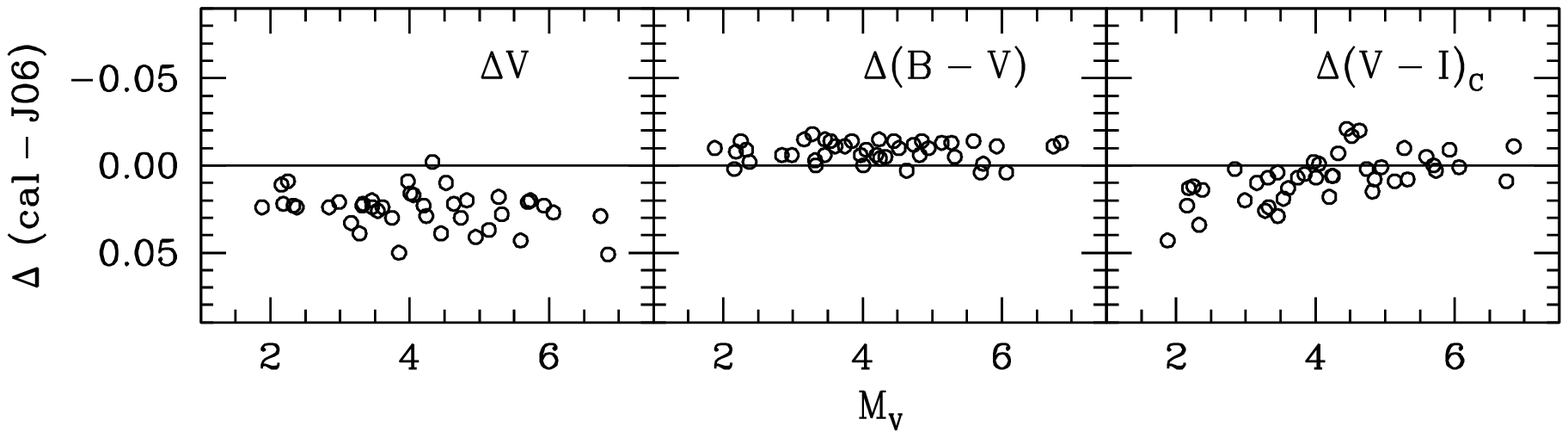}
\caption{Differences between the magnitudes and colors of the Hyades
members used to calibrate the Hyades empirical isochrone (Paper~II)
and photometry of \citet{joner06}. The differences are in the sense of
our compiled data minus Joner et~al. values. The solid line in each
panel indicates identity. The Johnson $V - I$ colors used in Paper~II
were transformed to the Cousins system using an updated transformation
(see text).\label{fig:joner}}
\end{figure}

Recently, \citet{joner06} published new Cousins photometry of a
large sample of Hyades stars. Their data set has the advantage over
the previous collection in that it was taken on a single telescope
and filter set. In Figure~\ref{fig:joner} we plot the
differences in the photometry for the stars used to calibrate the
Hyades isochrone (Paper~II) and the data in \citet{joner06}.
Compared to our values in Paper~II, their photometry is brighter in
$V$ and slightly redder in $B - V$. In $V - I_C$, the bright stars
in our sample are bluer than the Joner et al.\ values, but
both data are in agreement for the fainter stars. The mean
differences and standard errors for the range used in the MS fitting
of this paper ($M_V > 3.1$) are
$\Delta V = +0.026 \pm 0.002$, $\Delta(B - V) = -0.009 \pm 0.001$,
and $\Delta(V - I)_C = +0.004 \pm 0.002$.

These differences can be compared to those found previously
(Table~\ref{tab:phot.diff}) between the Tycho photometry and our values
in Paper~II. Compared to the Tycho-1, our Hyades values are also fainter
by 0.012 mag and are bluer in $B - V$ by 0.009 mag. If we had adjusted
zero points in $V$, $B - V$, and $V - I_C$ of isochrones to
the \citet{joner06} scales, the distances at spectroscopic metallicities
would become longer than our estimated values in Table~\ref{tab:dist} by
0.01 mag for Praesepe, 0.03 mag for the Pleiades, and 0.02 mag for both
M67 and NGC~2516.

\end{document}